\documentclass[reprint,twocolumn,superscriptaddress,amsmath,amssymb,aps,prb]{revtex4-2}
\usepackage{graphicx}
\usepackage{bm}
\usepackage{subcaption}
\usepackage{float}
\usepackage{lmodern}
\usepackage{mathtools}
\usepackage{placeins}
\usepackage[colorlinks=true,linkcolor=blue,urlcolor=blue,citecolor=blue,pdfusetitle]{hyperref}

\begin{document}

\title{Hubbard-assisted stability of Hatsugai--Kohmoto correlations\\in a one-dimensional open chain}

\author{Yan-Xiao Wang}
\affiliation{Key Laboratory of Quantum Theory and Applications of MoE \& School of Physical Science and Technology, Lanzhou University, Lanzhou 730000, China}
\affiliation{Lanzhou Center for Theoretical Physics, Key Laboratory of Theoretical Physics of Gansu Province, Lanzhou University, Lanzhou 730000, China}

\author{Yin Zhong}
\email{zhongy@lzu.edu.cn}
\affiliation{Key Laboratory of Quantum Theory and Applications of MoE \& School of Physical Science and Technology, Lanzhou University, Lanzhou 730000, China}
\affiliation{Lanzhou Center for Theoretical Physics, Key Laboratory of Theoretical Physics of Gansu Province, Lanzhou University, Lanzhou 730000, China}

\begin{abstract}
The Hatsugai-Kohmoto (HK) model has recently been identified as a fixed-point description of Mottness that is stable against perturbing local interactions. This raises a natural question beyond the weak-perturbation regime: how does a finite local repulsion modify HK physics in real-space observables and local spectra? We address this question in a half-filled one-dimensional spinful fermion chain, where the HK interaction is supplemented by an onsite Hubbard repulsion. Using exact diagonalization in an open chain, we compute ground-state correlation functions and site-resolved local spectra. We find that the Hubbard term does not drive the system away from the HK-dominated regime. Instead, at moderate $U_{\rm HK}$, a finite $U_{\rm Hub}$ promotes the emergence of behavior characteristic of the pure HK chain at larger $U_{\rm HK}$. This Hubbard-assisted strong-HK response is reflected consistently in the development of positive spin correlations, the suppression of single-particle coherence, the impurity-induced redistribution of local spectral weight, and the evolution of pairing correlations. As a control perturbation, we replace the onsite Hubbard interaction by a nearest-neighbor (NN) density interaction. In contrast to the Hubbard case, the HK and NN interactions display competing tendencies and do not reproduce the same strong-HK behavior. These results provide a nonperturbative real-space complement to the fixed-point stability of HK physics and show that the effect of an additional repulsive interaction depends sensitively on its spatial structure.
\end{abstract}

\maketitle

\section{Introduction}

The Hubbard model is a paradigmatic minimal model for correlated lattice electrons~\cite{Hubbard1963,Arovas2022}. With electron hopping and onsite repulsion, it captures the competition between itinerancy and local correlations and provides a common framework for discussing Mott physics, magnetism, and unconventional superconductivity~\cite{Mott1968,Gutzwiller1963,Kanamori1963,BrinkmanRice1970,Anderson1987,Dagotto1994,Georges1996,Imada1998,Essler2005,Lee2006}. Despite extensive progress, exact results are available only in special limits, such as the one-dimensional Bethe-ansatz solution~\cite{LiebWu1968,Essler2005} and the infinite-dimensional limit underlying dynamical mean-field theory~\cite{Metzner1989,Georges1996}. This has motivated the study of solvable or partially tractable models that retain essential aspects of strong correlation while exposing their many-body structure more directly~\cite{Baskaran1991,SachdevYe1993}.

One such example is the Hatsugai-Kohmoto (HK) model~\cite{HatsugaiKohmoto1992,ZhaoYangZhong2025Review}. The HK interaction is local in momentum space but nonlocal in real space, which makes the translationally invariant model exactly solvable while retaining Mott-like features. The resulting state is characterized by the redistribution of spectral weight and the appearance of zeros of the single-particle Green's function, providing a useful description of Mottness beyond the quasiparticle picture~\cite{Dzyaloshinskii2003,Phillips2020}. The HK framework has been used to study non-Fermi-liquid behavior, superconductivity, Friedel oscillations and impurity responses, spectral-weight transfer, transport, nonequilibrium dynamics, multiband and periodic-Anderson-type extensions, and the interplay between Mottness and topology~\cite{Lidsky1998,Yang2021,WangYang2023,Zhao2023,ZhaoYeoHuangPhillips2022,Zhong2022,Froldi2024,Mai2023,Tenkila2025,Guerci2025,Bacsi2025,ZhaoYangZhong2025Review}.

Recent work has further elevated the HK model from an exactly solvable construction to a fixed-point description of Mott physics. It has been proposed that the Mott transition can be understood as the breaking of a local-in-momentum-space $Z_2$ symmetry, with the HK interaction representing the Mott quartic fixed point that also governs the low-energy behavior of the Hubbard model~\cite{Huang2022}. A subsequent renormalization-group analysis further showed that this HK fixed point is stable against local interaction perturbations near the half-filled Mott regime, and that repulsive Hubbard interactions flow toward this stable fixed point~\cite{ZhaoLaNavePhillips2023}. These results suggest that HK physics is not merely a consequence of exact solvability, but rather a robust feature of strongly correlated electron systems.

These developments naturally raise a complementary question. Beyond the weak-perturbation regime addressed by renormalization-group arguments, how does a finite onsite Hubbard interaction modify HK correlations and impurity-induced local spectra? More importantly, is the resulting behavior unique to the onsite Hubbard interaction, or does it depend on the form of the added repulsive interaction? Such questions cannot be answered from the momentum-space fixed-point picture alone and instead require a direct real-space treatment in which the HK interaction, local perturbations, and finite geometry are treated on the same footing.

Most studies of HK physics exploit its momentum-space structure, where periodic boundary conditions (PBC) preserve translational invariance. The present work instead focuses on real-space correlations, impurity response, and site-resolved spectral functions, for which open boundary conditions (OBC) provide a natural setting. This choice is not merely a technical detail for the HK interaction, whose real-space form is long ranged. As emphasized in Ref.~\cite{Skolimowski2024}, OBC modify the allowed HK scattering processes because the center-of-mass constraint no longer includes processes that wrap around the chain as in periodic geometry. Consequently, boundary effects enter not only through the kinetic term but also through the HK interaction itself. An open-chain geometry therefore provides a natural framework for investigating real-space HK physics, while PBC are used only as a finite-size benchmark below.

In this work, we study a half-filled one-dimensional spinful chain with OBC using exact diagonalization in the real-space Fock basis. The central issue is how the real-space HK response is modified by an additional onsite Hubbard repulsion. The HK and Hubbard interactions represent two distinct forms of correlation: the former is momentum-local and real-space nonlocal, whereas the latter is strictly local in real space. If the fixed-point stability of HK physics has a nonperturbative real-space manifestation, one may expect that the Hubbard interaction reinforces rather than destroys the HK-dominated regime even beyond the weak-perturbation regime.

To address these questions, we compare the HK-only model, the Hubbard-only model, and the combined HK+Hubbard model within the same numerical framework. We examine equal-time spin and single-particle correlations, the local density of states (LDOS), impurity-induced LDOS evolution, and pairing correlations. To distinguish a genuine Hubbard-assisted effect from one arising simply from the presence of an additional repulsive interaction, we also consider a nearest-neighbor (NN) density-density interaction as a control perturbation.

Our main result is that the onsite Hubbard interaction reinforces the HK-dominated regime rather than producing an independent correlation pattern. At moderate $U_{\rm HK}$, increasing $U_{\rm Hub}$ drives the HK+Hubbard model toward behavior characteristic of the pure HK model at larger $U_{\rm HK}$. This Hubbard-assisted HK tendency is consistently reflected in the spin correlations, single-particle correlations, impurity-induced LDOS evolution, and pairing correlations. By contrast, an NN density interaction fails to reproduce this behavior and instead drives the system toward an NN-dominated response. These results provide a nonperturbative real-space manifestation of the fixed-point stability of HK physics and demonstrate that the effect of an additional repulsive interaction depends crucially on its spatial structure.

The remainder of the paper is organized as follows. In Sec.~\ref{sec:level2}, we introduce the model, observables, impurity setup, and numerical method. Sec.~\ref{sec:level3} presents the main results for the HK+Hubbard model. Sec.~\ref{sec:level4} examines an NN density interaction as a control perturbation. Finally, Sec.~\ref{sec:level5} summarizes the physical picture.

\section{Model and method}\label{sec:level2}

We consider a one-dimensional spinful fermion chain described by
\begin{equation}
 \begin{aligned}
  H=&-t\sum_{j,\sigma}\left(c^\dagger_{j\sigma}c_{j+1,\sigma}+{\rm H.c.}\right)-\mu\sum_{j,\sigma}n_{j\sigma} \\
  &+\frac{U_{\rm HK}}{L_x}\sum_{j_1,j_2,j_3,j_4}\delta_{j_1+j_3=j_2+j_4}c^\dagger_{j_1\uparrow}c_{j_2\uparrow}c^\dagger_{j_3\downarrow}c_{j_4\downarrow}\\
  &+U_{\rm Hub}\sum_j n_{j\uparrow}n_{j\downarrow}.
 \end{aligned}
\label{eq:ham}
\end{equation}

Here $c^\dagger_{j\sigma}$ creates a fermion with spin $\sigma=\uparrow,\downarrow$ on site $j=1,\ldots,L_x$ of a one-dimensional chain, where $L_x$ denotes the system size, $n_{j\sigma}=c^\dagger_{j\sigma}c_{j\sigma}$, and $t$ is taken as the unit of energy. The HK interaction has strength $U_{\rm HK}$, while $U_{\rm Hub}$ denotes the onsite Hubbard interaction. The constraint $\delta_{j_1+j_3=j_2+j_4}$ enforces conservation of the two-particle center-of-mass coordinate during HK scattering~\cite{HatsugaiKohmoto1992,Skolimowski2024}. Under OBC this is an ordinary equality of site indices, whereas under PBC it is understood modulo $L_x$.

We focus on the half-filled case, $N=N_\uparrow+N_\downarrow=L_x$. Unless otherwise specified, all calculations are performed for an $L_x=8$ open chain. PBC are considered only in Sec.~\ref{subsec:spin} for comparison with finite-size effects. Throughout this work we compare three cases: the HK-only model, the Hubbard-only model, and the combined HK+Hubbard model, corresponding to $(U_{\rm HK},U_{\rm Hub})=(U_{\rm HK},0)$, $(0,U_{\rm Hub})$, and finite values of both interactions, respectively.

The half-filled ground state is obtained directly in the fixed-$N$ sector without tuning the chemical potential. For the LDOS calculation, the chemical potential fixes the relative energy reference among the $N$, $N+1$, and $N-1$ sectors. We adopt the particle-hole symmetric convention
\begin{equation}
 \mu=\frac{U_{\rm HK}+U_{\rm Hub}}{2},
\label{eq:mu}
\end{equation}
which reduces to $\mu=U_{\rm HK}/2$ and $\mu=U_{\rm Hub}/2$ in the HK-only and Hubbard-only limits, respectively.

The LDOS for spin $\sigma$ on site $j$ is defined as
\begin{equation}
 A_{j\sigma}(\omega)=-\frac{1}{\pi}{\rm Im}\,G^R_{j\sigma}(\omega),
\label{eq:ldos}
\end{equation}
where $G^R_{j\sigma}(\omega)$ is the retarded local Green's function. When the half-filled ground state is degenerate, we employ the standard $T=0^+$ prescription, namely an equal-weight average over the degenerate ground-state manifold~\cite{HatsugaiKohmoto1992,Skolimowski2024,YeoPhillips2019}. For a ground-state degeneracy $g$, the Green's function is

\begin{equation}
 \begin{aligned}
 G^R_{j\sigma}(\omega)=\frac{1}{g}\sum_{a=1}^{g}\Bigg[&\sum_m\frac{\left|\langle m,N+1|c^\dagger_{j\sigma}|\Psi_a,N\rangle\right|^2}{\omega+i\eta-(E_m^{N+1}-E_0^N)}\\
 +&\sum_n\frac{\left|\langle n,N-1|c_{j\sigma}|\Psi_a,N\rangle\right|^2}{\omega+i\eta+(E_n^{N-1}-E_0^N)}\Bigg].
\end{aligned}
\label{eq:green}
\end{equation}

Here $|\Psi_a,N\rangle$ denotes one of the degenerate half-filled ground states, while $|m,N+1\rangle$ and $|n,N-1\rangle$ are many-body eigenstates in the particle-addition and particle-removal sectors with energies $E_m^{N+1}$ and $E_n^{N-1}$, respectively. A numerical broadening $\eta=0.02$ is used throughout the spectral calculations.

The LDOS is evaluated using the Lanczos continued-fraction method with $c^\dagger_{j\sigma}|\Psi_a,N\rangle$ and $c_{j\sigma}|\Psi_a,N\rangle$
as the initial vectors for the particle-addition and particle-removal parts, respectively~\cite{GaglianoBalseiro1987,Dagotto1994}. Since no spin-dependent perturbation, such as a Zeeman magnetic field, is present, the two spin components are equivalent, and only the $\sigma=\uparrow$ component is shown below.

We also calculate equal-time real-space correlation functions in the half-filled sector. For OBC, the reference site is fixed at the left edge, i.e., $j_0=1$. The spin correlation function is defined as
\begin{equation}
 C_S(\gamma)=\langle{\bf S}_{j_0+\gamma}\cdot{\bf S}_{j_0}\rangle,\qquad\gamma=0,\ldots,7,
\label{eq:spin_corr}
\end{equation}
where
\begin{equation}
 {\bf S}_j=\frac12\sum_{\sigma,\sigma'}c^\dagger_{j\sigma}{\boldsymbol\tau}_{\sigma\sigma'}c_{j\sigma'},
\label{eq:spin_operator}
\end{equation}
with ${\boldsymbol\tau}$ denoting the Pauli matrices. The single-particle correlation function is
\begin{equation}
 C_c(\gamma)=\langle c_{j_0+\gamma,\uparrow}c^\dagger_{j_0,\uparrow}\rangle,\qquad\gamma=0,\ldots,7.
\label{eq:single_particle_corr}
\end{equation}
When the ground state is degenerate, the same $T=0^+$ averaging procedure described above is adopted.

Pairing correlations are evaluated in the onsite singlet, NN singlet, and NN triplet channels. The corresponding pair operators are
\begin{equation}
 \Delta^{s_0}_i=c_{i\downarrow}c_{i\uparrow},
\label{eq:pair_s0}
\end{equation}

\begin{equation}
 \Delta^{s_1}_i=\frac{1}{\sqrt2}\left(c_{i\uparrow}c_{i+1,\downarrow}-c_{i\downarrow}c_{i+1,\uparrow}\right),
\label{eq:pair_s1}
\end{equation}
and
\begin{equation}
 \Delta^{t}_i=c_{i\uparrow}c_{i+1,\uparrow},
\label{eq:pair_t}
\end{equation}
where $s_0$, $s_1$, and $t$ denote the onsite singlet, NN singlet, and NN triplet channels, respectively. Their correlation functions are defined by
\begin{equation}
 P_\alpha(\gamma)=\left<\left(\Delta^\alpha_{j_0+\gamma}\right)^\dagger\Delta^\alpha_{j_0}\right>,\qquad\alpha=s_0,s_1,t,
\label{eq:pair_corr}
\end{equation}
using the same reference site $j_0=1$. For the $L_x=8$ open chain, $\gamma=0,\ldots,7$ for the onsite singlet channel and $\gamma=0,\ldots,6$ for the two NN bond channels.

To examine the response to a local perturbation, we introduce an impurity potential,
\begin{equation}
 H_{\rm imp}=V_{\rm imp}\left(n_{\ell\uparrow}+n_{\ell\downarrow}\right),
\label{eq:impurity}
\end{equation}
where $V_{\rm imp}$ is the impurity strength and $\ell$ denotes the impurity site. For the $L_x=8$ open chain, the impurity is placed at one of the two central sites, $\ell=L_x/2=4$, and the LDOS is evaluated on the neighboring site $j=5$.

As a control calculation, the onsite Hubbard interaction is replaced by a NN density-density interaction,
\begin{equation}
 H_V=V_{\rm NN}\sum_j n_j n_{j+1},\qquad n_j=n_{j\uparrow}+n_{j\downarrow},
\label{eq:nn_interaction}
\end{equation}
where $V_{\rm NN}$ denotes the interaction strength. The NN summation follows the same boundary convention as the hopping term. This interaction serves as a control perturbation for comparison with the onsite Hubbard interaction.

Unless otherwise specified, equal-time correlation functions are obtained by exact diagonalization in the real-space Fock basis, while the LDOS is evaluated using the Lanczos continued-fraction method~\cite{GaglianoBalseiro1987,Dagotto1994}.

\section{Results}\label{sec:level3}

We now turn to the numerical results. The HK-only and Hubbard-only limits reproduce the corresponding spin correlations, single-particle correlations, and local density of states reported in Ref.~\cite{Skolimowski2024}. These results therefore serve as benchmarks for the mixed HK+Hubbard system studied below. We first establish the spin-correlation signature of the strong-HK regime and then examine whether the same Hubbard-assisted trend is reflected in single-particle coherence, local spectra, pairing correlations, and impurity-induced spectral evolution.

\subsection{Spin correlations}\label{subsec:spin}

\begin{figure}[t]
\centering
\includegraphics[width=0.42\textwidth]{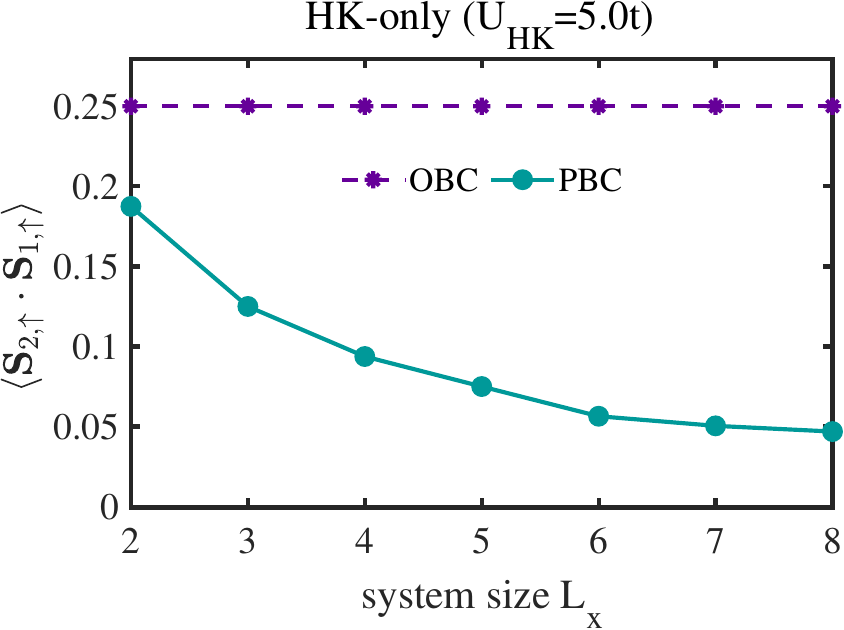}
\caption{Finite-size dependence of the NN spin correlation $C_S(1)$ in the HK-only model at $U_{\rm HK}=5t$, comparing OBC and PBC.}
\label{fig1}
\end{figure}

We begin with the finite-size behavior of the NN spin correlation $C_S(1)$, using it as a simple probe of the spin structure in the strong-HK regime. For the HK-only model at $U_{\rm HK}=5t$, Fig.~\ref{fig1} shows that the result depends strongly on the boundary condition. Under OBC, $C_S(1)$ remains very close to $1/4$ for all system sizes studied and shows no visible tendency to decrease with increasing $L_x$. Under PBC, in contrast, the same quantity is rapidly suppressed as the system size grows. Thus the positive NN spin correlation found in the open chain is not reproduced by the periodic finite-size sequence. This boundary-condition sensitivity is consistent with the real-space analysis of the HK interaction in Ref.~\cite{Skolimowski2024}, where OBC were shown to modify the allowed HK scattering processes through the center-of-mass constraint. It is therefore a characteristic feature of the strong-HK regime and cautions against treating OBC and PBC as interchangeable finite-size realizations.

\begin{figure}[H]
\centering
\includegraphics[width=0.425\textwidth]{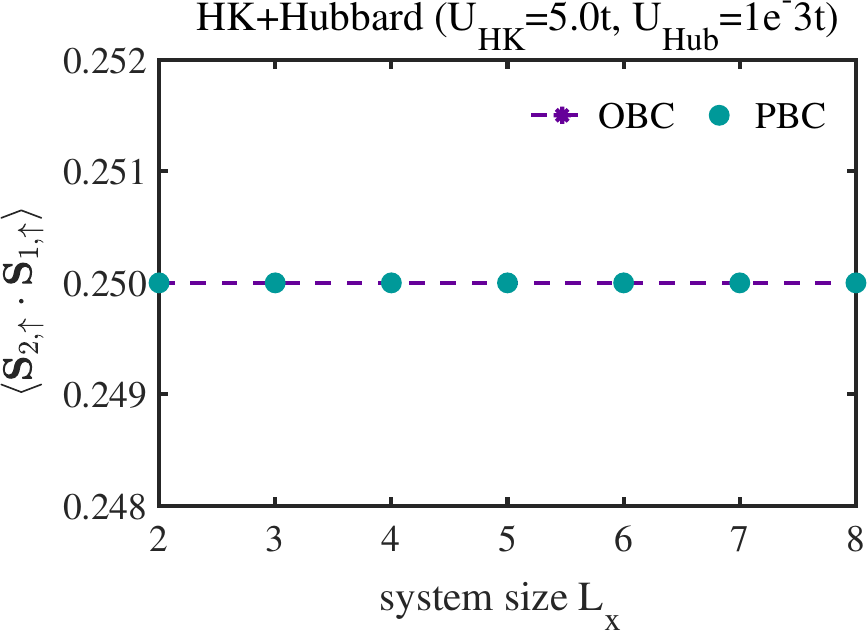}
\caption{Finite-size dependence of $C_S(1)$ in the HK+Hubbard model at $U_{\rm HK}=5t$ and $U_{\rm Hub}=10^{-3}t$, comparing OBC and PBC.}
\label{fig2}
\end{figure}

We next test whether this distinction survives once the HK model is perturbed by an onsite Hubbard repulsion. As shown in Fig.~\ref{fig2}, for $U_{\rm HK}=5t$ and $U_{\rm Hub}=10^{-3}t$, both OBC and PBC yield $C_S(1)\simeq 1/4$ over the accessible system sizes. Hence, an extremely weak Hubbard term already strongly reduces the OBC-PBC discrepancy of the pure HK limit and stabilizes the same positive NN spin correlation for the two boundary conditions. Together with the weak size dependence of the OBC data in Fig.~\ref{fig1}, this motivates the use of an $L_x=8$ open chain in the following real-space analysis, where the aim is to resolve the spatial structure of the correlations rather than to extract a thermodynamic critical point, which would require larger system sizes and careful finite-size extrapolation.

\begin{figure}[t]
\centering
\includegraphics[width=0.44\textwidth]{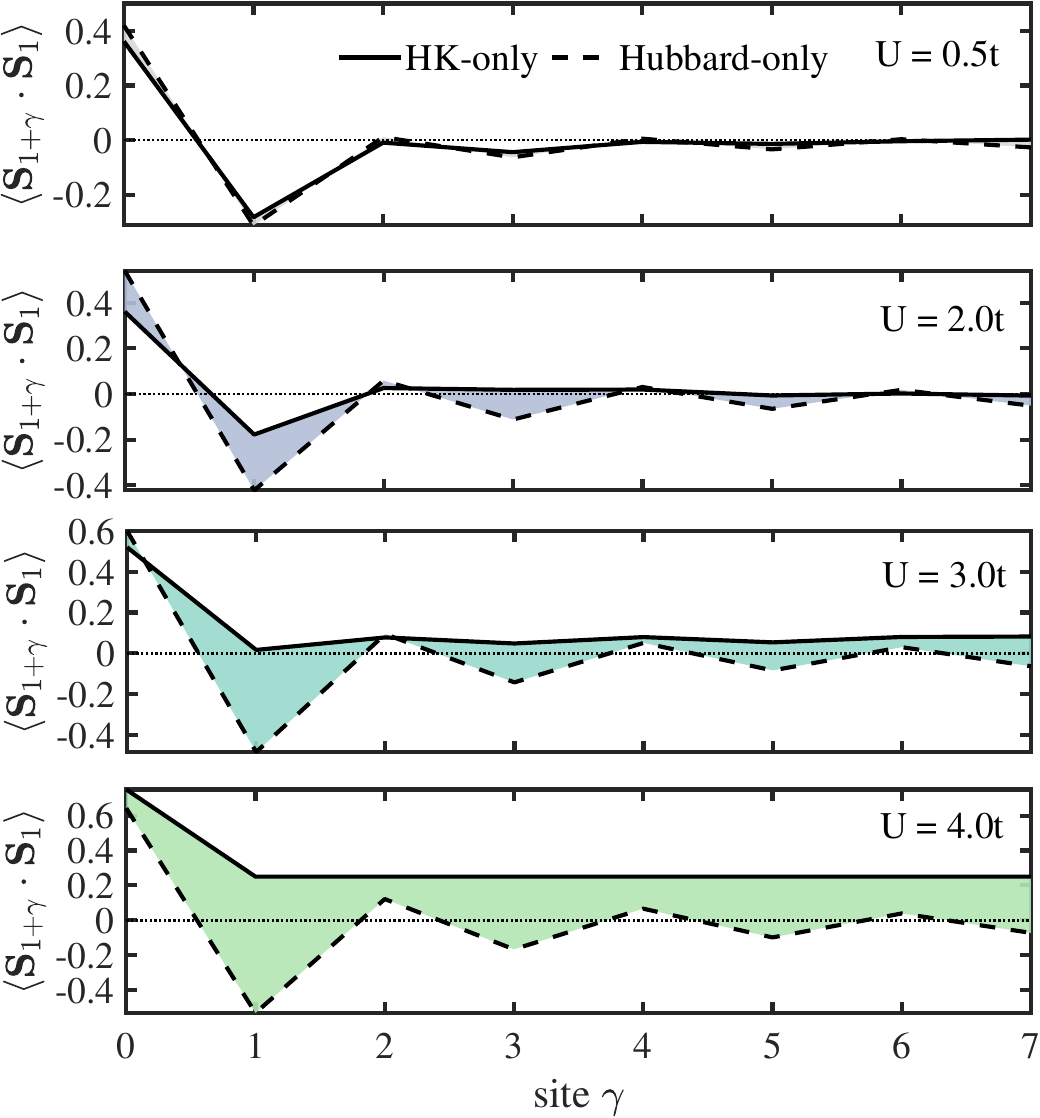}
\caption{Spin-spin correlation function $C_S(\gamma)$ for the HK-only and Hubbard-only models at selected interaction strengths. Solid lines denote the HK-only results with $U_{\rm HK}=U$, while dashed lines denote the Hubbard-only results with $U_{\rm Hub}=U$. The shaded regions highlight the difference between the two curves.}
\label{fig3}
\end{figure}

\begin{figure}[t]
\centering
\includegraphics[width=0.435\textwidth]{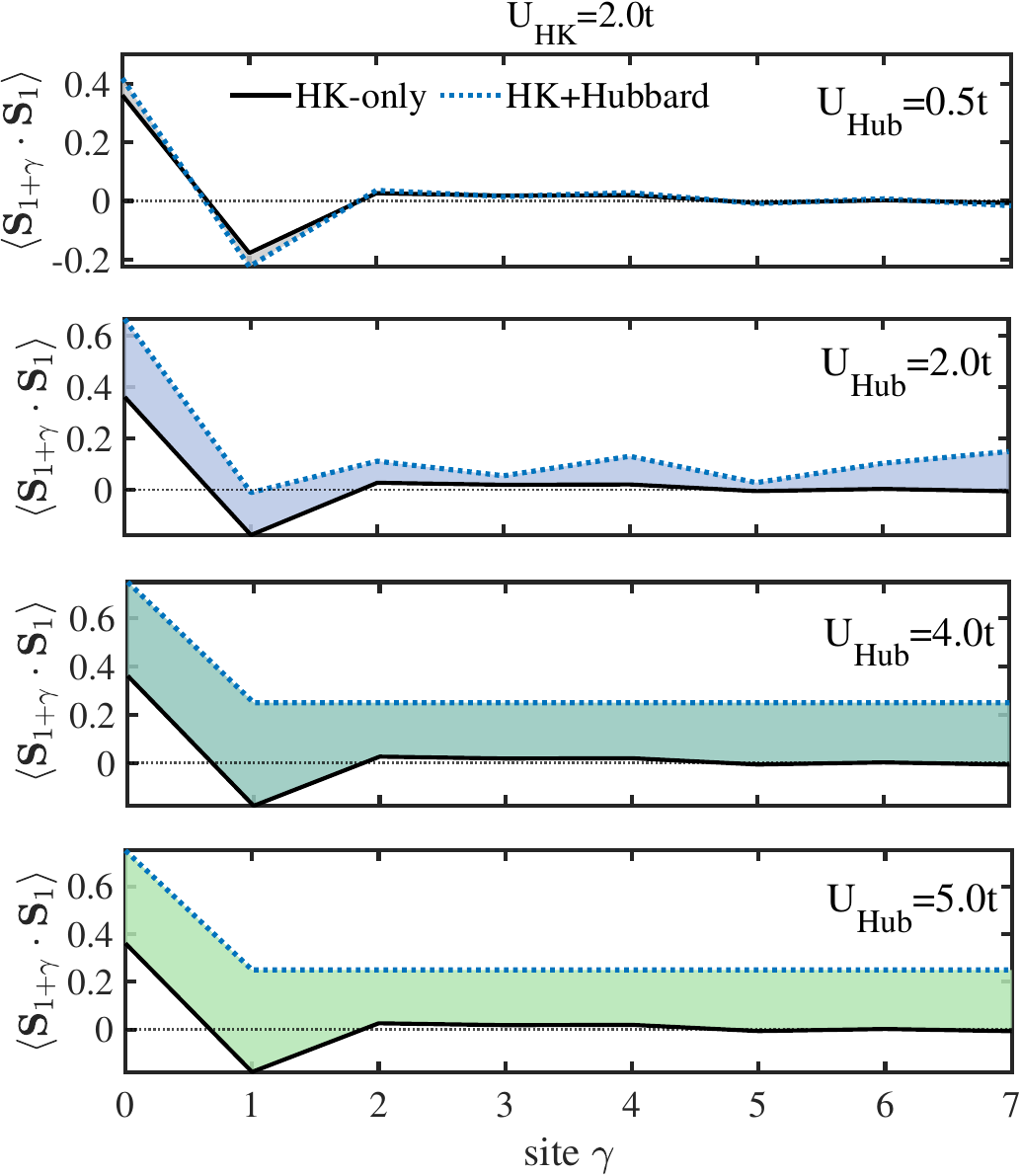}
\caption{$C_S(\gamma)$ at fixed $U_{\rm HK}=2t$. Solid black lines show the HK-only reference result, while blue dotted lines show the HK+Hubbard results for the indicated $U_{\rm Hub}$.}
\label{fig4}
\end{figure}

Having established the open-chain setting, we examine the full spatial profile of $C_S(\gamma)$. Fig.~\ref{fig3} compares the HK-only and Hubbard-only models at the same interaction strength. At weak coupling, the two models show similar short-range staggered correlations. Upon increasing $U$, however, their behavior separates clearly.

In the Hubbard-only model, $C_S(\gamma)$ remains oscillatory around zero: the NN component is negative, and the sign alternates with distance, as expected for antiferromagnetic correlations in the half-filled Hubbard chain. The amplitude changes with $U$, but the qualitative staggered structure is retained.

The HK-only model follows this Hubbard-like pattern only at small interaction strength. As $U_{\rm HK}$ is increased, the negative NN component is suppressed and the staggered structure gradually disappears. For $U_{\rm HK}=3t$, the intersite correlations have already turned positive over most of the chain. By $U_{\rm HK}=4t$, the intersite part of $C_S(\gamma)$ is nearly distance independent and remains close to $1/4$ for $\gamma\geq1$. This value has a simple two-spin interpretation: for two spin-$1/2$ moments, the operator $\mathbf{S}_i\cdot\mathbf{S}_j$ has eigenvalue $1/4$ in the triplet sector and $-3/4$ in the singlet sector. Therefore, an expectation value $\langle \mathbf{S}_i\cdot\mathbf{S}_j\rangle$ close to $1/4$ indicates that the corresponding spin correlation is near its ferromagnetic saturation value. The nearly saturated $C_S(\gamma)$ over the open chain thus supports an FM-like spin background in the strong-HK regime. This behavior is qualitatively different from the Hubbard-only reference, which keeps its staggered antiferromagnetic profile even at large $U_{\rm Hub}$.

\begin{figure}[H]
\centering
\includegraphics[width=0.46\textwidth]{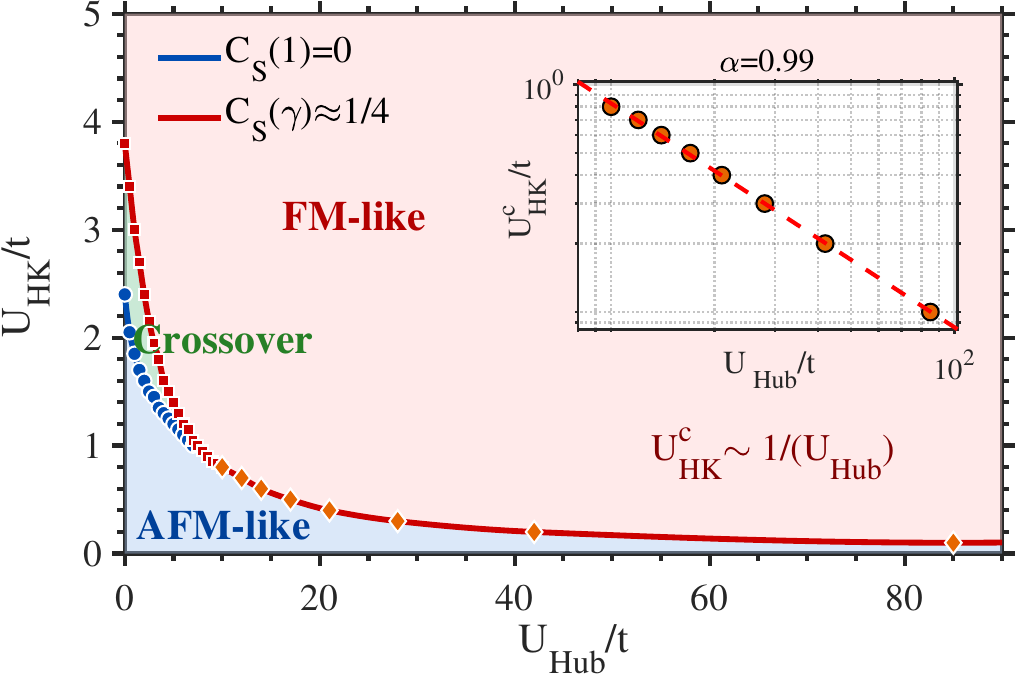}
\caption{Spin-correlation map in the $(U_{\rm Hub},U_{\rm HK})$ plane for the half-filled $L_x=8$ open chain. The blue line denotes the boundary defined by $C_S(1)=0$, beyond which the intersite spin correlations become positive. The red line denotes the boundary of the nearly saturated regime, where $C_S(\gamma)\simeq 1/4$ for $\gamma\geq 1$. Orange diamonds are extrapolated boundary values at large $U_{\rm Hub}$. The inset shows these extrapolated red-boundary values on a log-log scale, fitted by $U_{\rm HK}^c\propto (U_{\rm Hub})^{-\alpha}$ with $\alpha=0.99$. This fit is used only as an empirical guide to the large-$U_{\rm Hub}$ trend.}
\label{fig5}
\end{figure}

We now fix $U_{\rm HK}=2t$ and turn on the onsite Hubbard interaction. The results are shown in Fig.~\ref{fig4}. A small $U_{\rm Hub}$ only weakly modifies the HK-only curve. With increasing $U_{\rm Hub}$, however, the negative NN correlation is progressively removed, and the intersite correlations are lifted to positive values. For $U_{\rm Hub}=4t$ and $5t$, the profile becomes almost flat and approaches the same positive plateau found in the HK-only model at larger $U_{\rm HK}$. This is notable because, in this parameter range, the onsite Hubbard interaction is already larger than the HK interaction. Nevertheless, the mixed model does not evolve toward the large-$U_{\rm Hub}$ Hubbard-only pattern. Instead, the Hubbard term assists the HK interaction and drives the system toward the strong-HK spin pattern.

The broader parameter dependence is summarized in Fig.~\ref{fig5}. The blue line marks the sign change of the NN correlation, $C_S(1)=0$. On the high-$U_{\rm HK}$ side of this line, the intersite spin correlations become positive over the open chain. The red line marks the onset of the nearly saturated positive-correlation regime, where $C_S(\gamma)$ is close to $1/4$ for $\gamma\geq1$. Increasing $U_{\rm Hub}$ shifts this red boundary rapidly toward smaller $U_{\rm HK}$, showing that the onsite Hubbard repulsion substantially reduces the HK interaction strength required to reach the strong-HK spin pattern. The orange diamonds denote extrapolated red-boundary values at large $U_{\rm Hub}$. As shown in the inset, these points are fitted on a log-log scale by $U_{\rm HK}^c\propto (U_{\rm Hub})^{-\alpha}$ with $\alpha=0.99$, indicating an empirical inverse-like trend of the boundary at large $U_{\rm Hub}$.

These results identify the spin channel as the clearest manifestation of the Hubbard-assisted HK tendency. In the absence of the HK interaction, the Hubbard repulsion enhances spin correlations but preserves the conventional staggered profile. Once the HK interaction is present, however, the same onsite repulsion no longer produces a Hubbard-only pattern. Instead, it promotes the positive, nearly distance-independent spin correlations characteristic of the strong-HK regime. Since the present calculations are performed on finite chains, the boundaries in Fig.~\ref{fig5} should be understood as operational indicators of changes in the real-space correlation pattern, rather than as thermodynamic phase boundaries.

\subsection{Single-particle correlations}\label{subsec:single}

After analyzing the spin correlations, we examine whether similar trends appear in single-particle motion. The single-particle correlation function $C_c(\gamma)$ probes the spatial coherence between two sites separated by a distance $\gamma$.

\begin{figure}[t]
\centering
\includegraphics[width=0.42\textwidth]{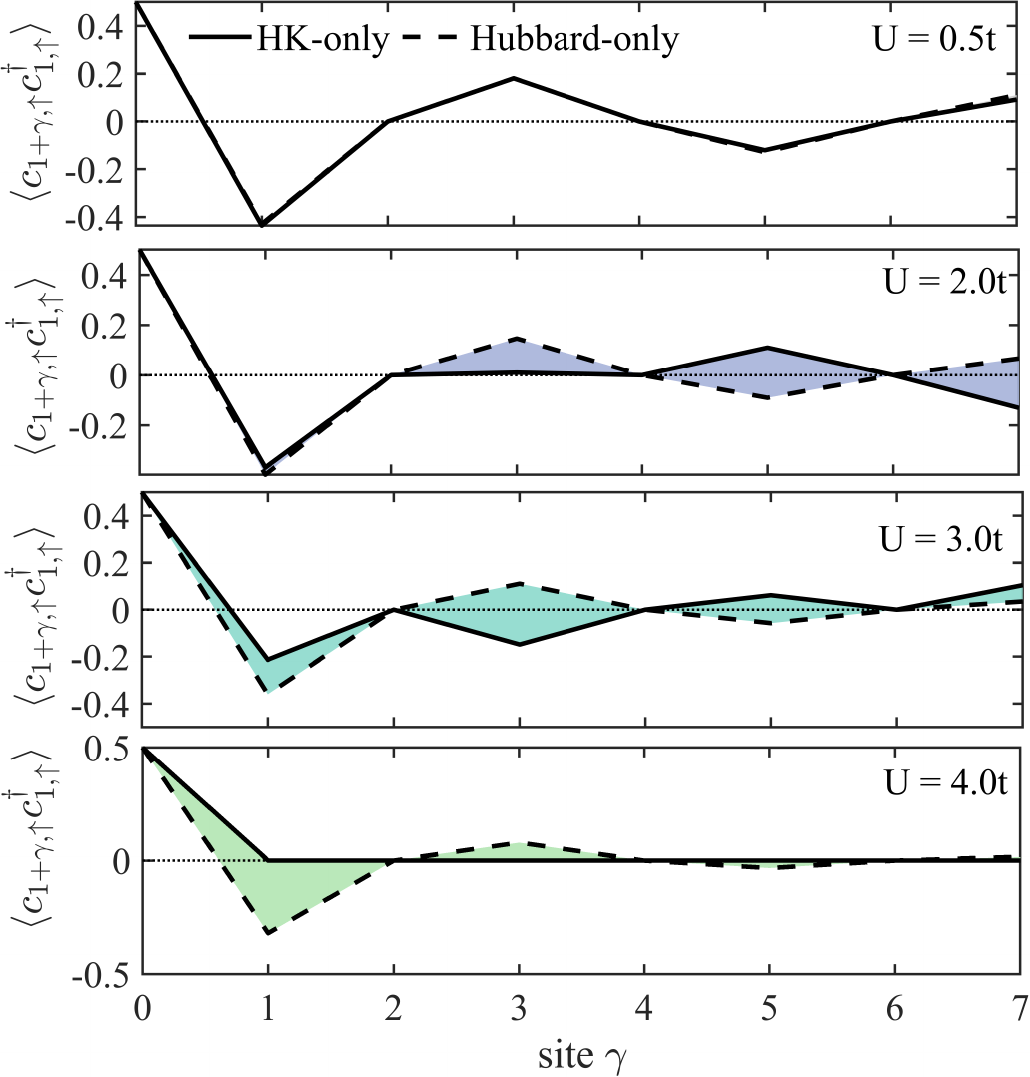}
\caption{Single-particle correlation function $C_c(\gamma)$ for the HK-only (solid lines) and Hubbard-only (dashed lines) models at selected interaction strengths.}
\label{fig6}
\end{figure}

Fig.~\ref{fig6} compares the HK-only and Hubbard-only models at the same interaction strength, with $U_{\rm HK}=U$ and $U_{\rm Hub}=U$, respectively. At weak coupling ($U=0.5t$), the two curves are nearly identical, reflecting the similarity of the weak-coupling ground states in the two models. Upon increasing $U$, however, their nonlocal correlations evolve in clearly different ways. The Hubbard-only model retains an oscillatory tail over several lattice spacings, whereas the HK-only model suppresses intersite coherence much more efficiently, consistent with the strong real-space reorganization induced by the HK interaction. At large $U$, the HK-only correlation is strongly suppressed away from the reference site, indicating a rapid loss of single-particle coherence.

This rapid damping of $C_c(\gamma)$ is a characteristic feature of the strong-HK regime. However, it should not be interpreted as defining a sharp boundary analogous to the magnetic crossover discussed in Sec.~\ref{subsec:spin}. In the intermediate interaction regime, where spin correlations have already lost their staggered structure, single-particle coherence can still extend over a significant portion of the finite chain. Therefore, $C_c(\gamma)$ does not provide an independent criterion for the magnetic boundary; instead, it reflects a complementary aspect of the same HK-dominated physics.

From a physical perspective, enhanced parallel-spin correlations reduce the available phase space for coherent single-particle propagation due to Pauli constraints. This naturally leads to a faster suppression of fermionic coherence in the HK-only model compared with the purely local Hubbard model, even though both contain strong repulsive interactions.

\begin{figure}[t]
\centering
\includegraphics[width=0.425\textwidth]{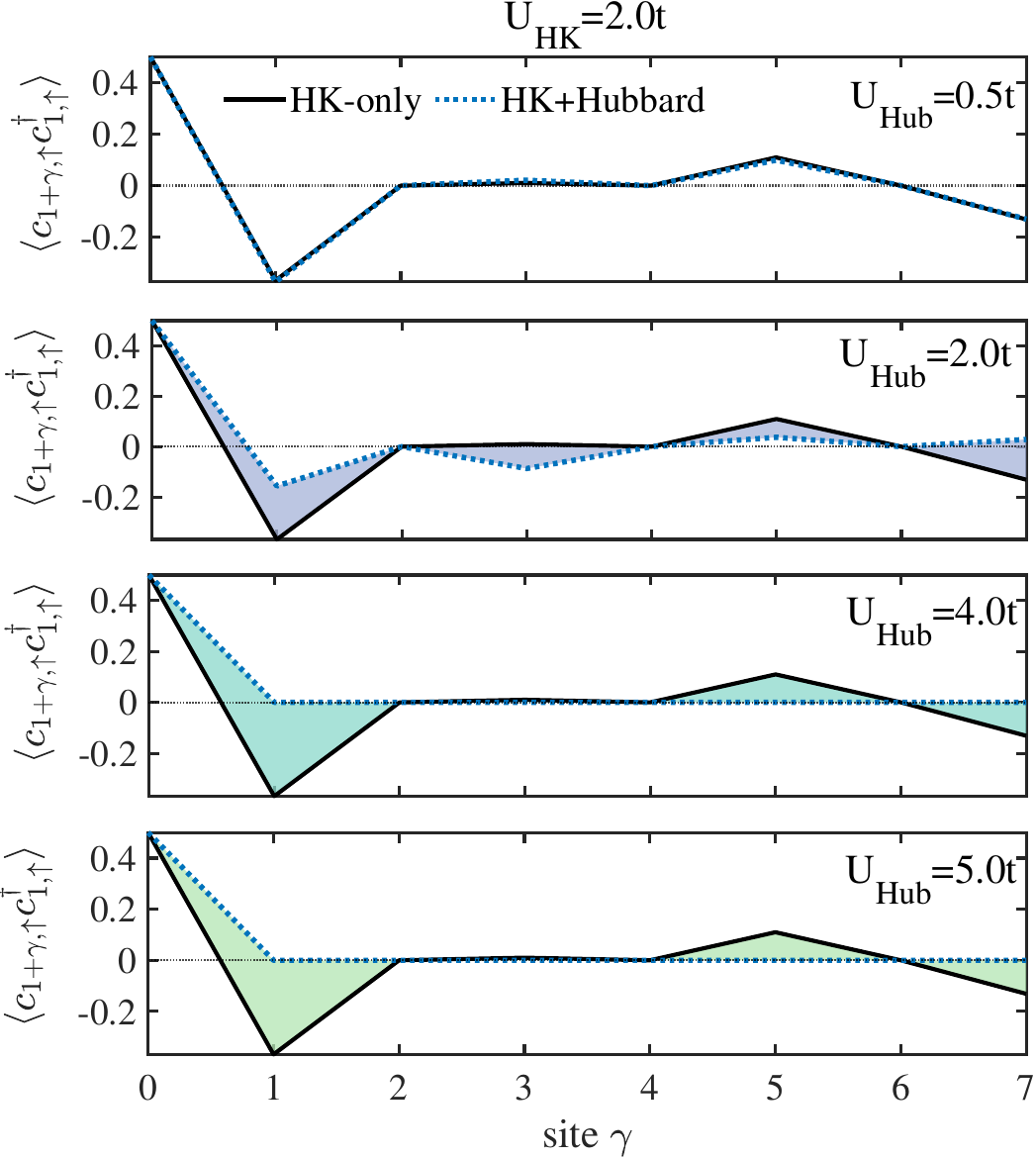}
\caption{$C_c(\gamma)$ at fixed $U_{\rm HK}=2t$. Solid black lines show the HK-only reference result, while blue dotted lines show the HK+Hubbard results for the indicated $U_{\rm Hub}$.}
\label{fig7}
\end{figure}

The effect of adding an onsite Hubbard term to the HK model is shown in Fig.~\ref{fig7}, where $U_{\rm HK}$ is fixed at $2t$. A weak Hubbard interaction produces only minor corrections to the HK-only reference curve. For larger $U_{\rm Hub}$, however, the nonlocal part of $C_c(\gamma)$ is strongly suppressed. In particular, for $U_{\rm Hub}=4t$ and $5t$, correlations away from the reference site are almost completely flattened.

This behavior shows that the Hubbard term does not drive the system toward a distinct Hubbard-like oscillatory regime. Instead, it reinforces the HK tendency and pushes the mixed system toward the strong-HK limit, here manifested as a loss of fermionic coherence. The single-particle correlations therefore provide a complementary signature of the Hubbard-assisted enhancement of HK physics.

\subsection{Local density of states}
\label{subsec:ldos}

The equal-time correlations show how the Hubbard term modifies the static real-space structure. We next examine the corresponding local spectral signatures. For the $L_x=8$ open chain, the spectrum is evaluated at site $j=5$ in the spin-up channel and is denoted by $A_{5,\uparrow}(\omega)$. 

\begin{figure}[t]
\centering
\includegraphics[width=0.44\textwidth]{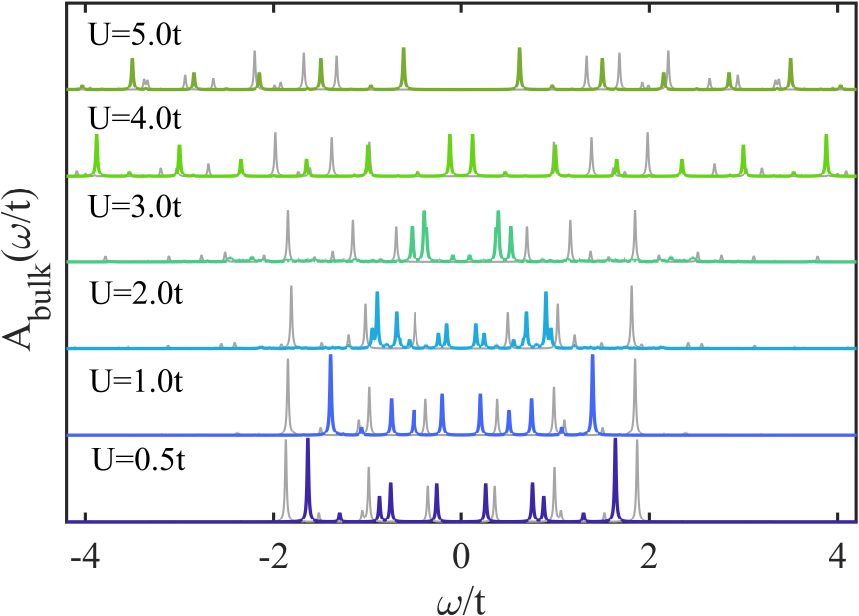}
\caption{LDOS $A_{5,\uparrow}(\omega)$ at site $j=5$ for the HK-only and Hubbard-only models at selected interaction strengths. Colored and gray curves denote the HK-only and Hubbard-only results, respectively. The spectra are vertically shifted for clarity.}
\label{fig8}
\end{figure}

Fig.~\ref{fig8} compares the HK-only and Hubbard-only models at the same interaction strength. At weak coupling, the two spectra have similar low-energy structures. With increasing interaction, however, the dominant peaks evolve in different ways. In the Hubbard-only model, the main low-energy peaks move steadily away from $\omega=0$, reflecting the increasing onsite charge energy. In the HK-only model, the lowest dominant peaks first approach $\omega=0$ and then separate again at stronger $U_{\rm HK}$. This nonmonotonic evolution is a characteristic feature of the HK spectrum and is not described by a simple Hubbard-band splitting.

The weaker high-energy structures also distinguish the two limits. In the Hubbard-only model, spectral weight is transferred to higher energies already at moderate $U_{\rm Hub}$, and the corresponding weak peaks become more pronounced as the interaction grows. In the HK-only model, these additional features develop more slowly, while the strongest peaks remain tied to the low-energy structure over a wider range of $U_{\rm HK}$. The HK interaction therefore redistributes spectral weight in a manner different from that produced by a purely local Hubbard repulsion.

\begin{figure}[t]
\centering
\includegraphics[width=0.44\textwidth]{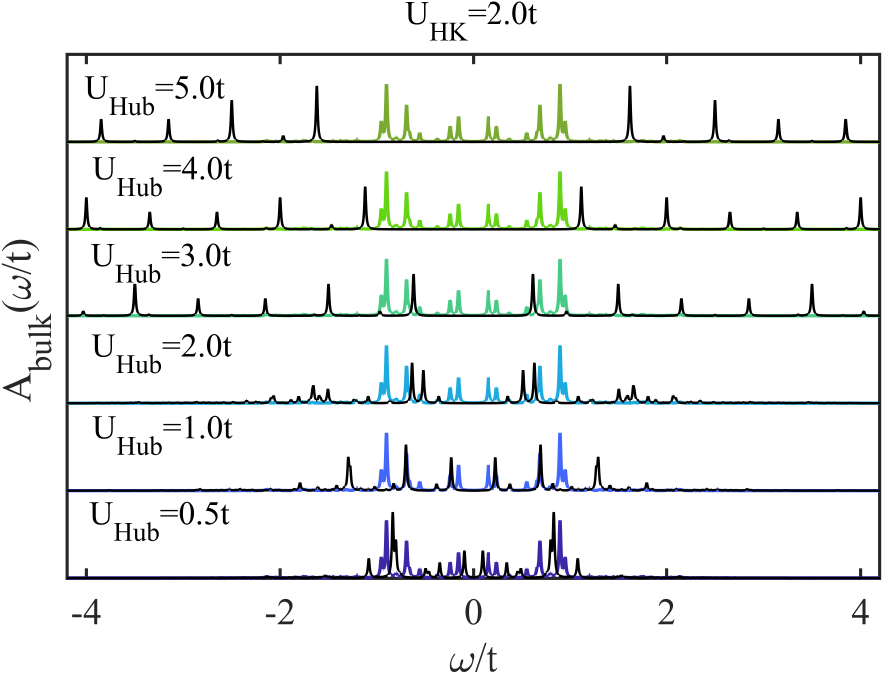}
\caption{LDOS $A_{5,\uparrow}(\omega)$ at fixed $U_{\rm HK}=2t$ for several values of $U_{\rm Hub}$. Colored curves denote the HK-only reference spectra, while black curves denote the HK+Hubbard results. }
\label{fig9}
\end{figure}

Fig.~\ref{fig9} shows the effect of adding the onsite Hubbard interaction at fixed $U_{\rm HK}=2t$. For small $U_{\rm Hub}$, the spectrum remains close to the HK-only reference. As $U_{\rm Hub}$ is increased, the main particle-addition and particle-removal peaks move apart, indicating that the Hubbard term introduces an additional local charge energy scale. This change, however, should not be interpreted as a conversion into the Hubbard-only spectrum. The peak fragmentation and the redistribution of spectral weight retain the structure characteristic of the HK interaction.

The LDOS thus supports the picture obtained from the equal-time correlations. The Hubbard term adds a local charge scale, but it does not generate an independent Hubbard-like spectral pattern. Instead, the low-energy peaks separate while the overall spectral structure remains governed by the HK-induced redistribution of spectral weight.

\subsection{Pairing correlations}
\label{subsec:pairing}

We also examine the onsite singlet, NN singlet, and NN triplet pairing correlations defined in Sec.~\ref{sec:level2}. This provides a complementary check of whether the Hubbard-assisted HK regime contains an additional pairing tendency.

\begin{figure}[H]
\centering
\includegraphics[width=0.49\textwidth]{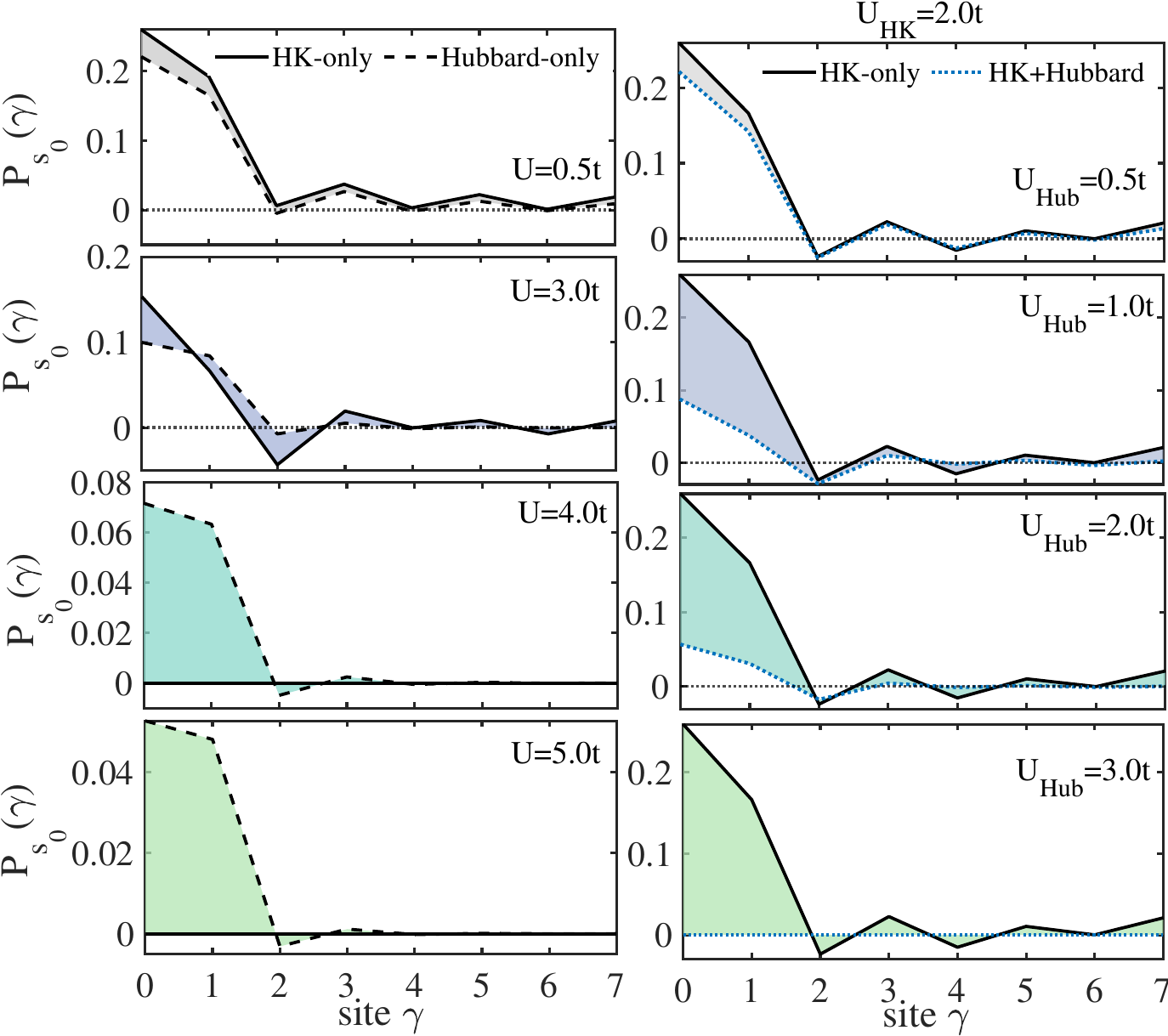}
\caption{Onsite singlet pairing correlation $P_{s_0}(\gamma)$. Left column: comparison between the HK-only (solid lines) and Hubbard-only (dashed lines) models at selected interaction strengths. Right column: results at fixed $U_{\rm HK}=2t$ for several values of $U_{\rm Hub}$. The solid black lines denote the HK-only reference, while the blue dotted lines denote the HK+Hubbard model.}
\label{fig10}
\end{figure}

The onsite singlet channel is shown in Fig.~\ref{fig10}. At weak interaction, the HK-only and Hubbard-only results are nearly identical. With increasing interaction strength, the HK interaction suppresses $P_{s_0}(\gamma)$ over the whole chain, whereas the Hubbard-only model retains finite short-range weight at $\gamma=0$ and $\gamma=1$. At fixed $U_{\rm HK}=2t$, increasing $U_{\rm Hub}$ further suppresses the onsite singlet correlation, bringing the mixed model closer to the strong-HK limit.

\begin{figure}[t]
\centering
\includegraphics[width=0.49\textwidth]{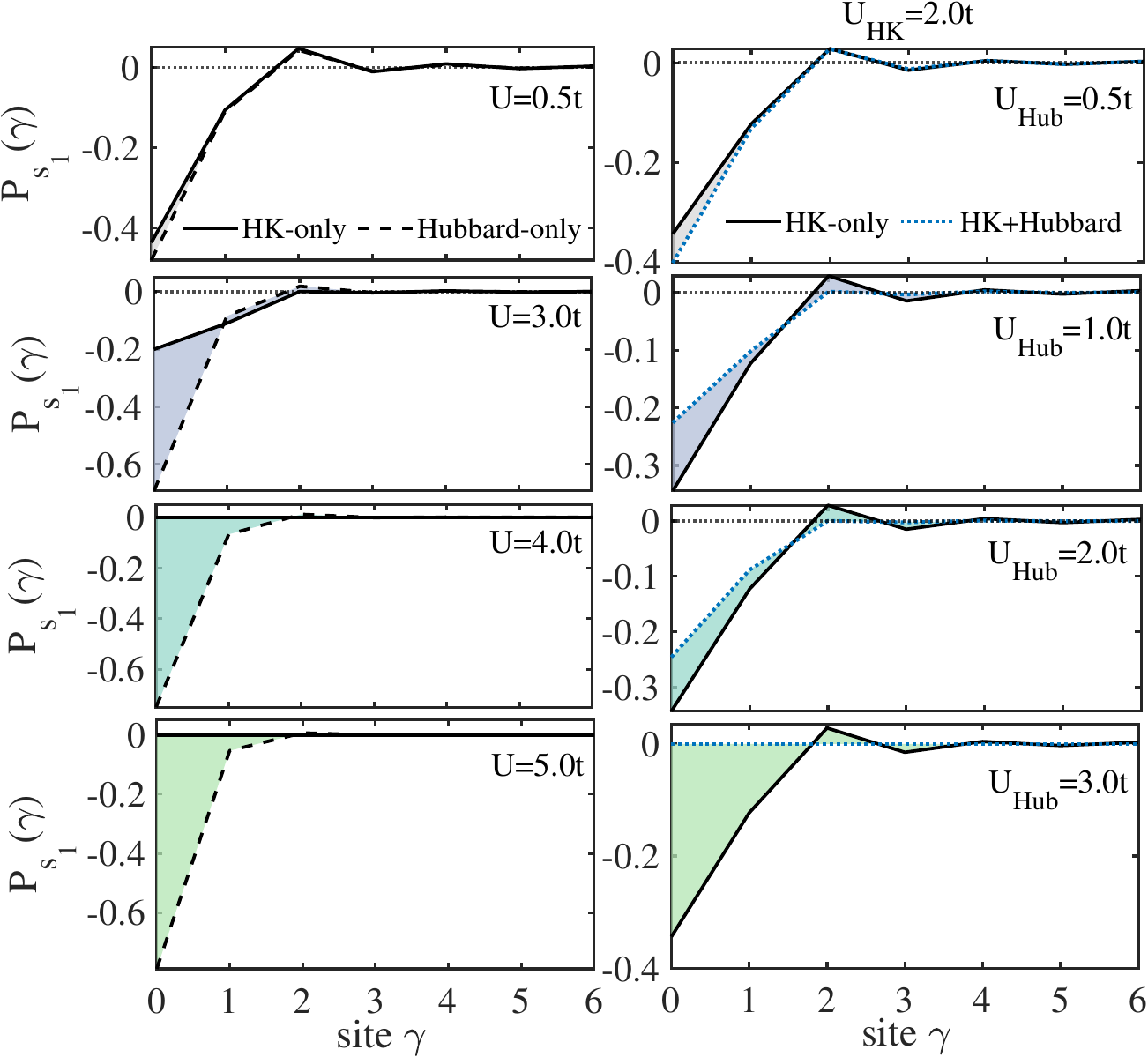}
\caption{NN singlet pairing correlation $P_{s_1}(\gamma)$. Left column: comparison between the HK-only (solid lines) and Hubbard-only (dashed lines) models at selected interaction strengths. Right column: results at fixed $U_{\rm HK}=2t$ for several values of $U_{\rm Hub}$. The solid black lines denote the HK-only reference, while the blue dotted lines denote the HK+Hubbard model.}
\label{fig11}
\end{figure}

The NN singlet channel, shown in Fig.~\ref{fig11}, displays its main difference at the nearest-bond component $P_{s_1}(0)$. At weak interaction, the two pure limits are almost indistinguishable. At larger interaction strengths, $P_{s_1}(0)$ remains sizable and negative in the Hubbard-only model, while its magnitude is strongly reduced in the HK-only model and approaches zero. Longer-distance components are small in both cases. At fixed $U_{\rm HK}=2t$, increasing $U_{\rm Hub}$ reduces the magnitude of the NN singlet correlation in the mixed model. Thus, the Hubbard term does not enhance NN singlet pairing, but drives this channel toward the strong-HK profile.

\begin{figure}[t]
\centering
\includegraphics[width=0.49\textwidth]{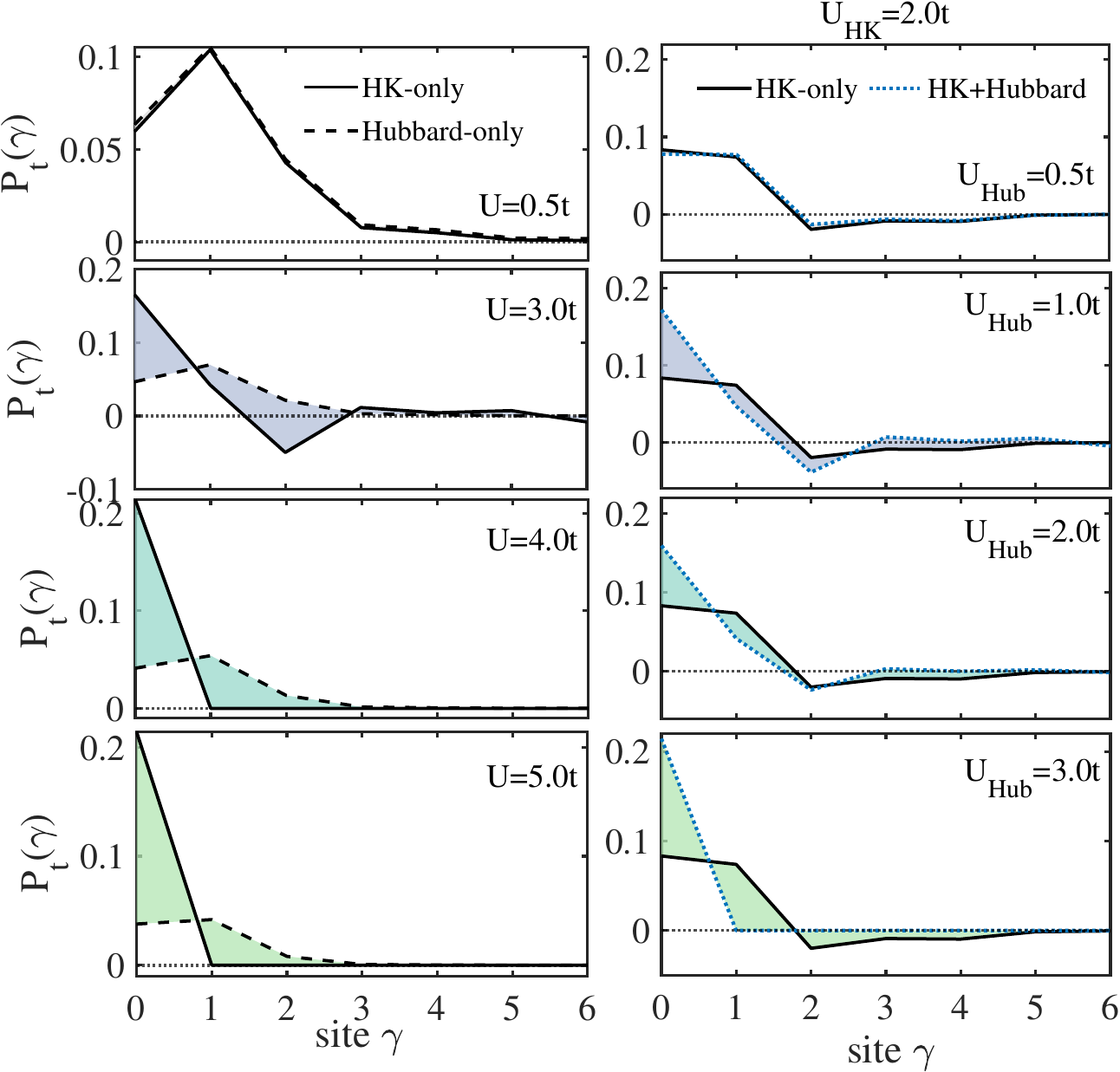}
\caption{NN equal-spin triplet pairing correlation $P_t(\gamma)$. Left column: comparison between the HK-only (solid lines) and Hubbard-only (dashed lines) models at selected interaction strengths. Right column: results at fixed $U_{\rm HK}=2t$ for several values of $U_{\rm Hub}$. The solid black lines denote the HK-only reference, while the blue dotted lines denote the HK+Hubbard model.}
\label{fig12}
\end{figure}

We next consider the NN triplet channel in Fig.~\ref{fig12}. Unlike the singlet channels, the strong-HK behavior is characterized by a redistribution of the triplet weight toward the shortest bond. In the HK-only model, $P_t(\gamma)$ is concentrated mainly at $\gamma=0$, while finite-distance correlations are strongly suppressed. The Hubbard-only model instead retains weak short-range correlations around $\gamma=1$. At fixed $U_{\rm HK}=2t$, increasing $U_{\rm Hub}$ enhances the $\gamma=0$ component and suppresses the remaining finite-distance weight, again driving the mixed model toward the HK-only profile.

Taken together, the pairing correlations do not indicate an independent Hubbard-induced pairing channel. The onsite and NN singlet components are suppressed, while the NN triplet component becomes localized on the shortest bond, consistent with the strong-HK correlation pattern identified above.

\begin{figure}[H]
\centering
\includegraphics[width=0.46\textwidth]{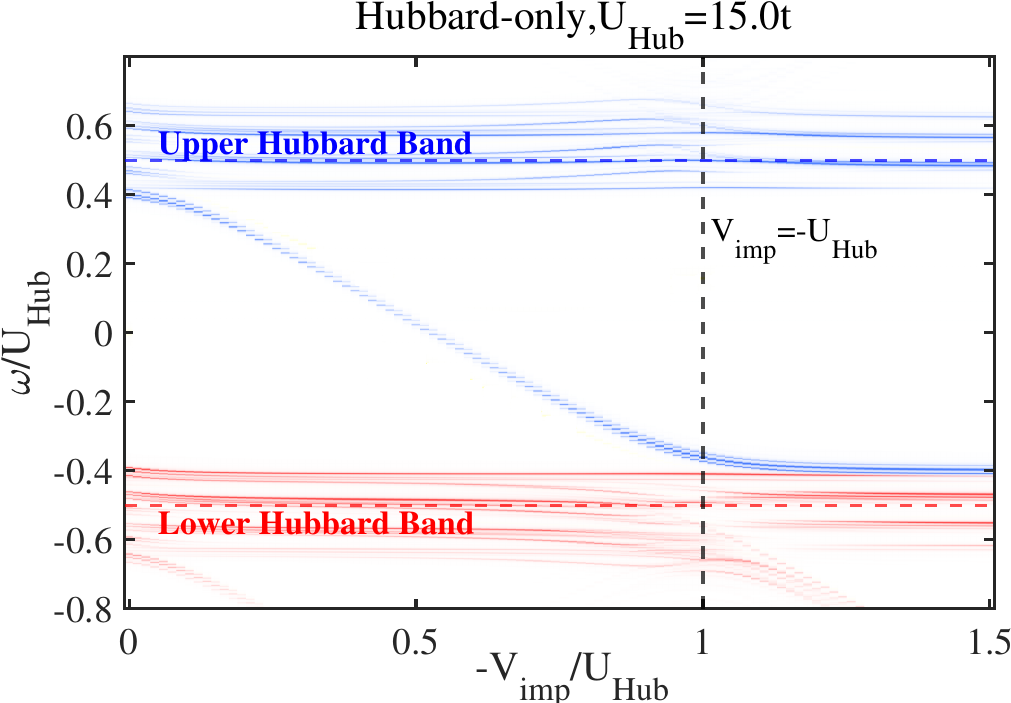}
\caption{Impurity-dependent local spectrum $A_{5,\uparrow}(\omega)$ of the Hubbard-only model at $U_{\rm Hub}=15t$. Blue and red branches denote particle addition and removal, respectively; the dashed line marks $V_{\rm imp}=-U_{\rm Hub}$.}
\label{fig14}
\end{figure}

\begin{figure*}[t]
\centering
\includegraphics[width=0.9\textwidth]{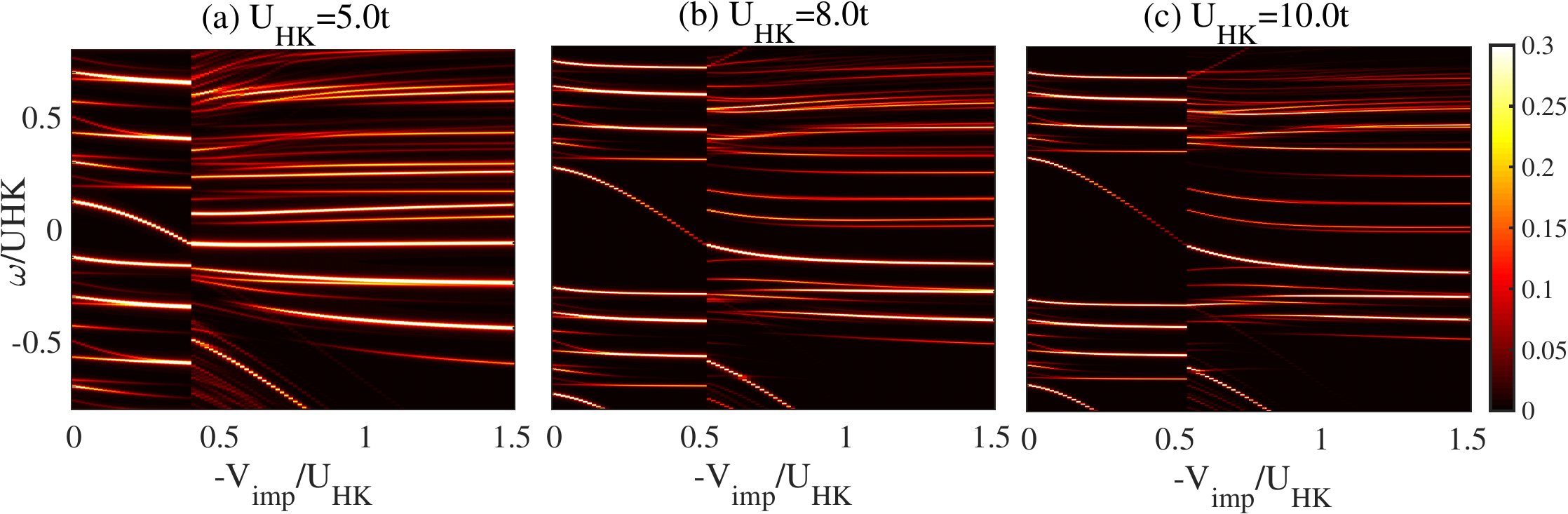}
\caption{$A_{5,\uparrow}(\omega)$ in the HK-only model for (a) $U_{\rm HK}=5t$, (b) $8t$, and (c) $10t$. The dominant impurity-induced resonance approaches the low-energy region and tends to saturate near $\omega\simeq0$ as $|V_{\rm imp}|$ is increased.}
\label{fig15}
\end{figure*}

\subsection{Impurity-induced LDOS evolution}
\label{subsec:impurity_ldos}

We now examine the response to a local potential in Eq.~\eqref{eq:impurity}. Impurity spectroscopy has recently been proposed as a useful probe of Green's-function zeros and correlated Mott spectral structures, since impurity-induced in-gap spectral weight can reveal information beyond that contained in quasiparticle poles alone~\cite{Mitra2026}. Motivated by this idea, we compare the impurity-induced LDOS in the Hubbard-only, HK-only, and HK+Hubbard systems.

We first consider the Hubbard-only reference system at strong coupling, $U_{\rm Hub}=15t$, shown in Fig.~\ref{fig14}. An attractive impurity induces a particle-addition resonance that starts from the bottom of the upper Hubbard band. As $|V_{\rm imp}|$ increases, this resonance is continuously pulled across the Mott gap and approaches the top of the lower Hubbard band. This is the conventional impurity response of a Mott insulator, where the local potential transfers spectral weight between the two Hubbard bands.

The HK-only model shows a different impurity response, as displayed in Fig.~\ref{fig15}. An attractive potential again produces a downward-moving resonance, but its evolution is not a simple transfer from one Hubbard band edge to the other. Instead, the dominant impurity-induced branch enters the correlated gap and gradually becomes concentrated in the low-energy region. For large $|V_{\rm imp}|$, its position changes only weakly and tends to saturate near $\omega\simeq0$, i.e., close to the Fermi level.

This behavior is consistent with the local impurity $T$-matrix picture. For a single-site potential, the resonance condition can be written as
\[
1-V_{\rm imp}G^R_{\ell\ell}(\omega)=0 ,
\]
or equivalently $G^R_{\ell\ell}(\omega)\simeq1/V_{\rm imp}$. In the strong-scattering limit $|V_{\rm imp}|\rightarrow\infty$, the resonance is therefore driven toward energies where the local Green's function is small. In the HK system, the correlated gap is associated with strongly suppressed low-energy spectral weight and a zero-like structure of the Green's function near the Fermi level. The impurity-induced branch is therefore attracted toward this low-energy structure and tends to accumulate near $\omega\simeq0$, rather than continuing to move toward a Hubbard-band edge.

Several weaker branches are also visible in the HK spectra. These features reflect the finite-size level structure and the rearrangement of HK many-body states under the impurity potential. Their presence does not originate from the Lanczos continued-fraction procedure, as checked by comparison with full diagonalization in Appendix~\ref{app:hk_discontinuity}.

\begin{figure}[t]
\centering
\includegraphics[width=0.485\textwidth]{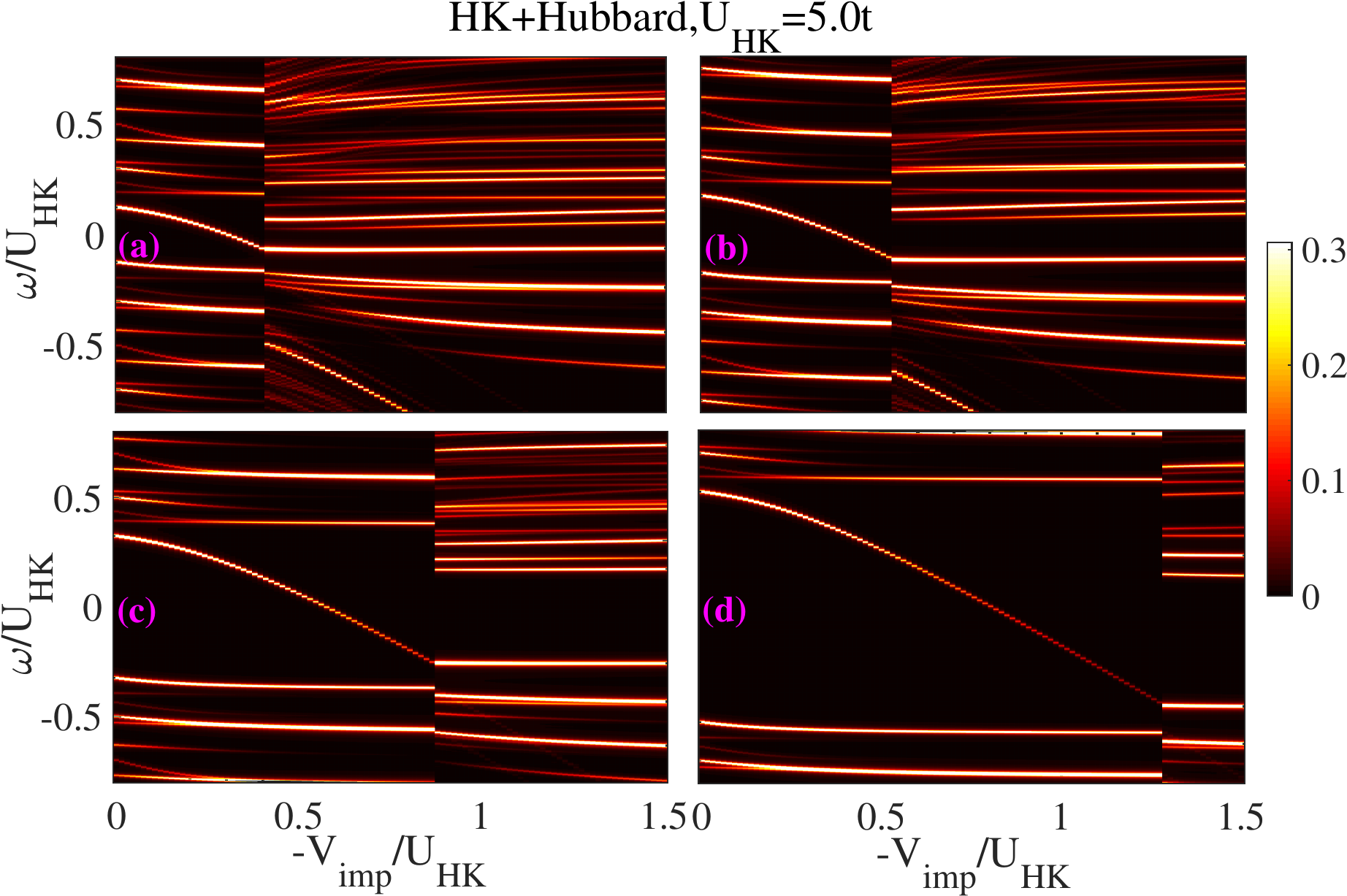}
\caption{$A_{5,\uparrow}(\omega)$ at fixed $U_{\rm HK}=5t$ for (a) $U_{\rm Hub}=0$, (b) $0.5t$, (c) $2t$, and (d) $4t$.}
\label{fig16}
\end{figure}

\begin{figure*}[t]
\centering
\includegraphics[width=0.9\textwidth]{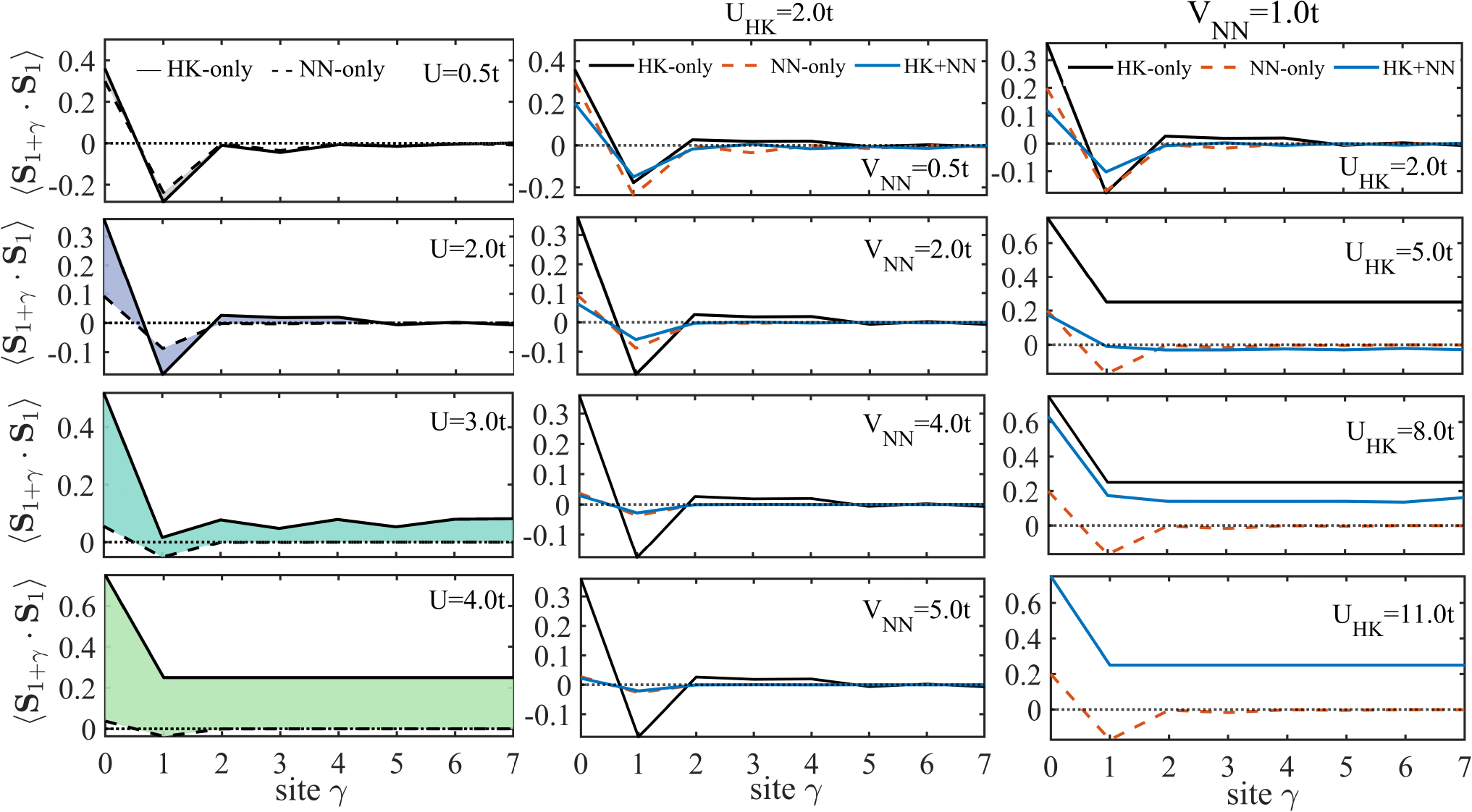}
\caption{Spin correlation function $C_S(\gamma)$ in the HK-only, NN-only, and HK+$V_{\rm NN}$ models. Left: comparison at equal interaction magnitude. Middle: fixed $U_{\rm HK}=2t$ and varying $V_{\rm NN}$. Right: fixed $V_{\rm NN}=t$ and varying $U_{\rm HK}$.}
\label{fig17}
\end{figure*}

We finally add the onsite Hubbard interaction to the HK model at fixed $U_{\rm HK}=5t$, as shown in Fig.~\ref{fig16}. The Hubbard term does not restore the conventional band-to-band impurity response seen in the Hubbard-only limit. Instead, the spectrum retains the HK-type in-gap resonance structure. With increasing $U_{\rm Hub}$, the dominant impurity-induced branch becomes more isolated from the higher-energy background and shifts to negative energies. Thus, the Hubbard interaction reorganizes the in-gap spectral weight of the HK impurity response, rather than pinning the resonance more strongly at $\omega=0$.

Overall, the impurity LDOS distinguishes two types of behavior. In the Hubbard-only Mott regime, the impurity resonance connects the upper and lower Hubbard bands. In the HK-dominated regime, the impurity response is governed by in-gap spectral redistribution: in the HK-only case the dominant resonance tends to saturate near the Fermi level, while in the HK+Hubbard case the in-gap branch remains pronounced but is shifted away from $\omega=0$ by the onsite repulsion.

\section{Nearest-neighbor density interaction as a control perturbation}
\label{sec:level4}

We now use the nearest-neighbor density interaction in Eq.~\eqref{eq:nn_interaction} as a control perturbation. The purpose is not to determine the full phase diagram of the HK+$V_{\rm NN}$ model, but to test whether the Hubbard-assisted HK behavior found above is a generic consequence of adding a repulsive interaction. We therefore compare the HK-only, NN-only, and mixed HK+$V_{\rm NN}$ models using the same real-space observables as in Sec.~\ref{sec:level3}.

Fig.~\ref{fig17} compares the spin correlations. In the left column, the HK-only and NN-only models are shown at the same interaction strength. The two limits are clearly distinct. With increasing interaction, the HK-only model develops sizable positive spin correlations over the chain, whereas the NN-only result stays close to zero except at the shortest distances. The middle column fixes $U_{\rm HK}=2t$ and increases $V_{\rm NN}$. In this case, the added NN density interaction suppresses the HK-like positive component and drives the mixed result toward the NN-only profile. Conversely, when $V_{\rm NN}=t$ is fixed and $U_{\rm HK}$ is increased, the positive HK-like spin correlation is gradually recovered. Thus, unlike the HK+Hubbard case, the HK+$V_{\rm NN}$ model does not show a cooperative enhancement of the strong-HK spin pattern. Its spin correlations are instead controlled mainly by the competition between the two interaction scales.

\begin{figure*}[t]
\centering
\includegraphics[width=0.9\textwidth]{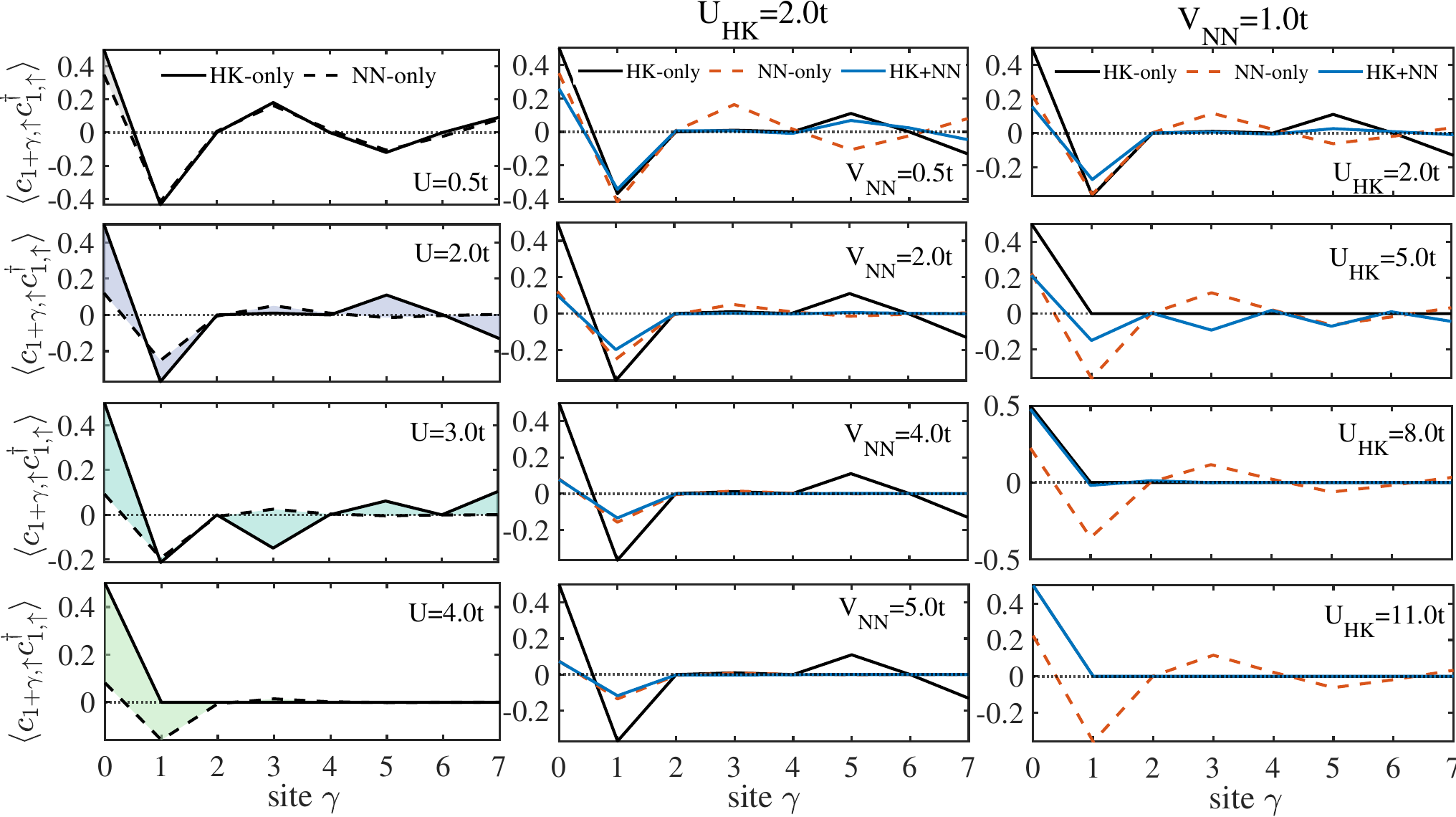}
\caption{Single-particle correlation function $C_c(\gamma)$ in the same three comparisons as Fig.~\ref{fig17}.}
\label{fig18}
\end{figure*}

The single-particle correlations in Fig.~\ref{fig18} show the same contrast. At equal interaction strength, the HK-only and NN-only limits have different spatial structures. At fixed $U_{\rm HK}=2t$, increasing $V_{\rm NN}$ weakens the HK-dominated profile and brings the mixed result closer to the NN-only behavior. When $V_{\rm NN}=t$ is fixed, increasing $U_{\rm HK}$ restores the HK-like suppression of finite-distance single-particle coherence. The NN density interaction therefore does not generate an assisted HK regime in the single-particle channel.

These control results show that the assisted behavior found for the onsite Hubbard term is not a generic consequence of adding an additional repulsive interaction to the HK model. The onsite Hubbard interaction reinforces the HK tendency and pushes the mixed model toward the behavior of a pure HK chain at stronger $U_{\rm HK}$. The nearest-neighbor density interaction behaves differently: it mainly competes with the HK interaction and suppresses the HK-like signatures until the HK scale becomes dominant.

\section{Conclusion}\label{sec:level5}

We have studied how onsite and nearest-neighbor repulsive interactions modify the half-filled one-dimensional HK model in real space. Motivated by the proposed stability of the HK fixed point against local perturbations~\cite{Huang2022,ZhaoLaNavePhillips2023}, we explored a regime beyond the weak-perturbation limit addressed by renormalization-group arguments. Using exact diagonalization on open chains, we examined how finite short-range interactions affect the real-space signatures of HK physics.

Our results reveal a Hubbard-assisted enhancement of HK-type correlations in finite systems. Rather than interpolating smoothly between the HK-only and Hubbard-only limits, the onsite Hubbard interaction at moderate $U_{\rm HK}$ drives the system toward correlation patterns characteristic of stronger HK coupling. This indicates that, in the parameter regime studied here, onsite repulsion reinforces rather than competes with the HK-induced correlation tendency.

The enhancement is most clearly manifested in the spin sector. In the pure HK model, increasing $U_{\rm HK}$ generates positive and nearly saturated spin correlations. At fixed $U_{\rm HK}$, adding a finite $U_{\rm Hub}$ produces the same qualitative evolution and lowers the interaction strength required to reach the strong-HK spin pattern. Finite-size comparisons further show that even a weak Hubbard term substantially reduces the OBC--PBC discrepancy present in the pure HK limit, indicating that onsite repulsion stabilizes the real-space spin response.

The other observables show consistent but channel-dependent behavior. Single-particle correlations exhibit enhanced suppression of spatial coherence with increasing $U_{\rm Hub}$. In the impurity LDOS, the additional Hubbard interaction does not restore the conventional Hubbard-band spectral transfer; instead, the in-gap spectral redistribution characteristic of the HK regime remains pronounced, with the detailed resonance evolution modified by the onsite repulsion. Pairing correlations are suppressed in both onsite and NN singlet channels, while the equal-spin triplet component retains the residual short-range structure associated with the strong-HK regime.

In contrast, replacing the onsite Hubbard interaction by a nearest-neighbor density interaction does not produce the same assisted behavior. The HK+$V_{\rm NN}$ model instead exhibits a competition between HK-like and NN-dominated correlations, with the dominant response controlled by the relative interaction strengths. This comparison shows that the Hubbard-assisted enhancement is not a generic consequence of adding repulsive interactions, but is closely related to the onsite structure of the Hubbard term.

We further find that this Hubbard-assisted HK tendency survives away from half filling, although doping modifies the detailed forms of the spin and single-particle correlation functions. The evolution toward stronger-HK-like correlation patterns remains visible over a finite doping range.

Overall, these results provide a nonperturbative real-space perspective on the stability and enhancement of HK physics beyond weak-coupling arguments. While previous RG analyses established the robustness of the HK fixed point against weak local perturbations, our results show that onsite interactions can actively reinforce HK-like correlations in finite systems, whereas spatially extended density interactions tend to compete with them. This distinction highlights the importance of interaction geometry in determining whether HK correlations are enhanced or suppressed.

\begin{acknowledgments}
This work was supported by the National Natural Science Foundation of China (Grant No.~12247101), the Fundamental Research Funds for the Central Universities (Grant No.~lzujbky-2024-jdzx06), the Natural Science Foundation of Gansu Province (Grant Nos.~22JR5RA389 and 25JRRA799), and the National ``111 Center'' for Collaborative Research (Grant No.~B20063).
\end{acknowledgments}

\appendix

\section{Benchmark of the HK impurity spectrum}
\label{app:hk_discontinuity}

In the HK-only impurity spectra, several weak branches show abrupt rearrangements as the impurity potential is varied. Such discontinuous spectral features are not unexpected in the HK model, since the exact translationally invariant solution already exhibits discontinuities in the spectral weight associated with the highly degenerate mixed-state ground-state manifold~\cite{Skolimowski2024}. In the present finite open chain with a local impurity, varying $V_{\rm imp}$ can similarly reorganize the nearly degenerate many-body levels and redistribute weak spectral weight. We therefore benchmark the result against full diagonalization to verify that these structures are not artifacts of the Lanczos continued-fraction procedure.

\begin{figure}[H]
\vspace{1.5em}
\centering
\includegraphics[width=0.48\textwidth]{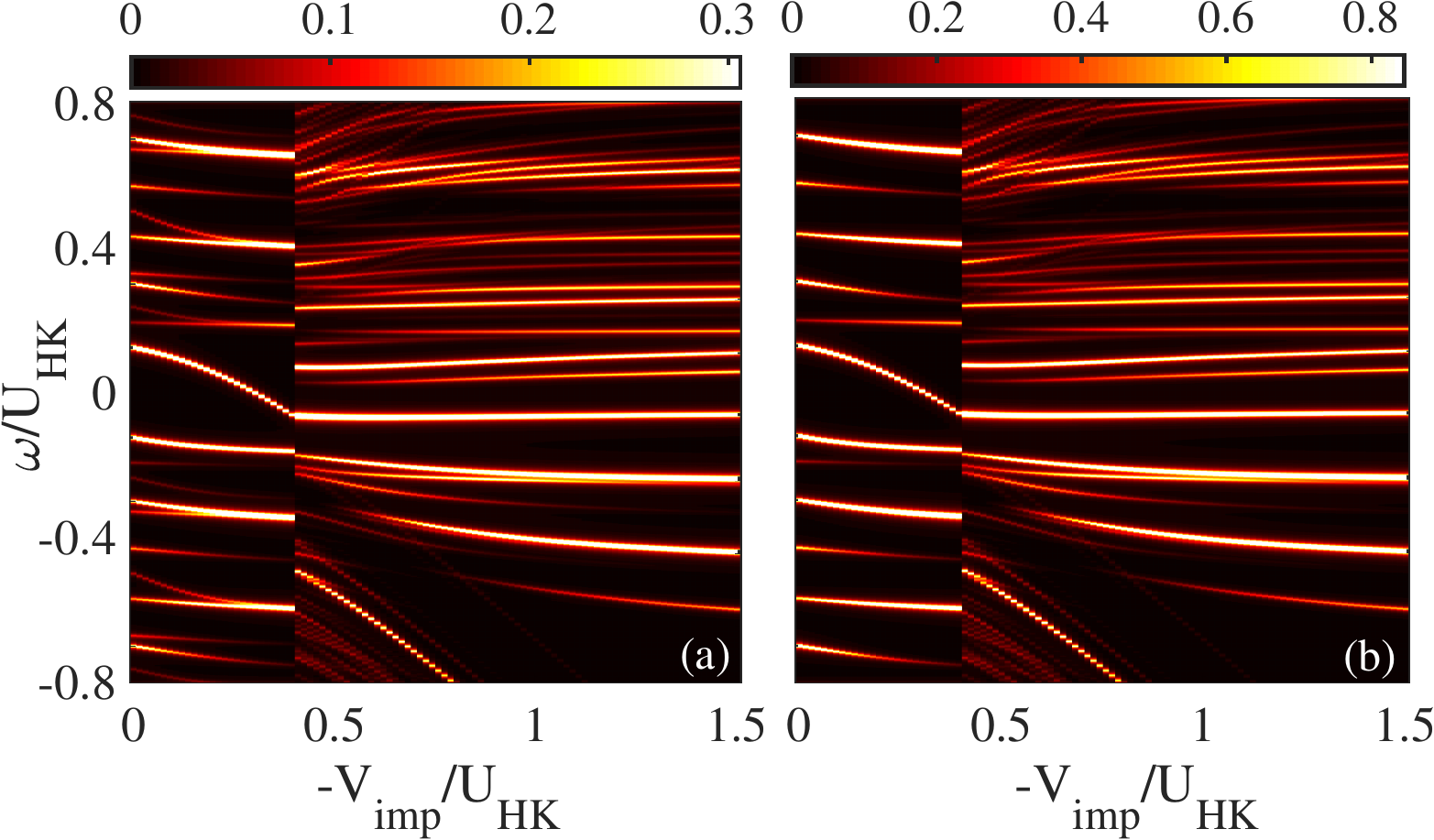}
\caption{Local density of states $A_{5,\uparrow}(\omega)$ of the pure HK model at $U_{\rm HK}/t=5$, obtained from (a) the Lanczos continued-fraction method and (b) full diagonalization with averaging over the degenerate ground-state manifold. The two methods give nearly identical spectral structures; the color scales are set independently.}
\label{fig19}
\end{figure}

Fig.~\ref{fig19} compares the two calculations for the pure HK model at $U_{\rm HK}/t=5$. They produce nearly identical spectral branches, including the weak discontinuous features. This agreement confirms that the discontinuities are intrinsic to the finite-size HK spectrum rather than artifacts of the Lanczos implementation. The different color ranges only affect the displayed peak intensities and do not change the branch structure. In the main text, we therefore focus on the dominant impurity-induced resonance, whose evolution is robust in both calculations.

\section{Correlations away from half filling}
\label{app:away_half_filling}

\begin{figure*}[t]
\vspace{1.5em}
\centering
\includegraphics[width=0.8\textwidth]{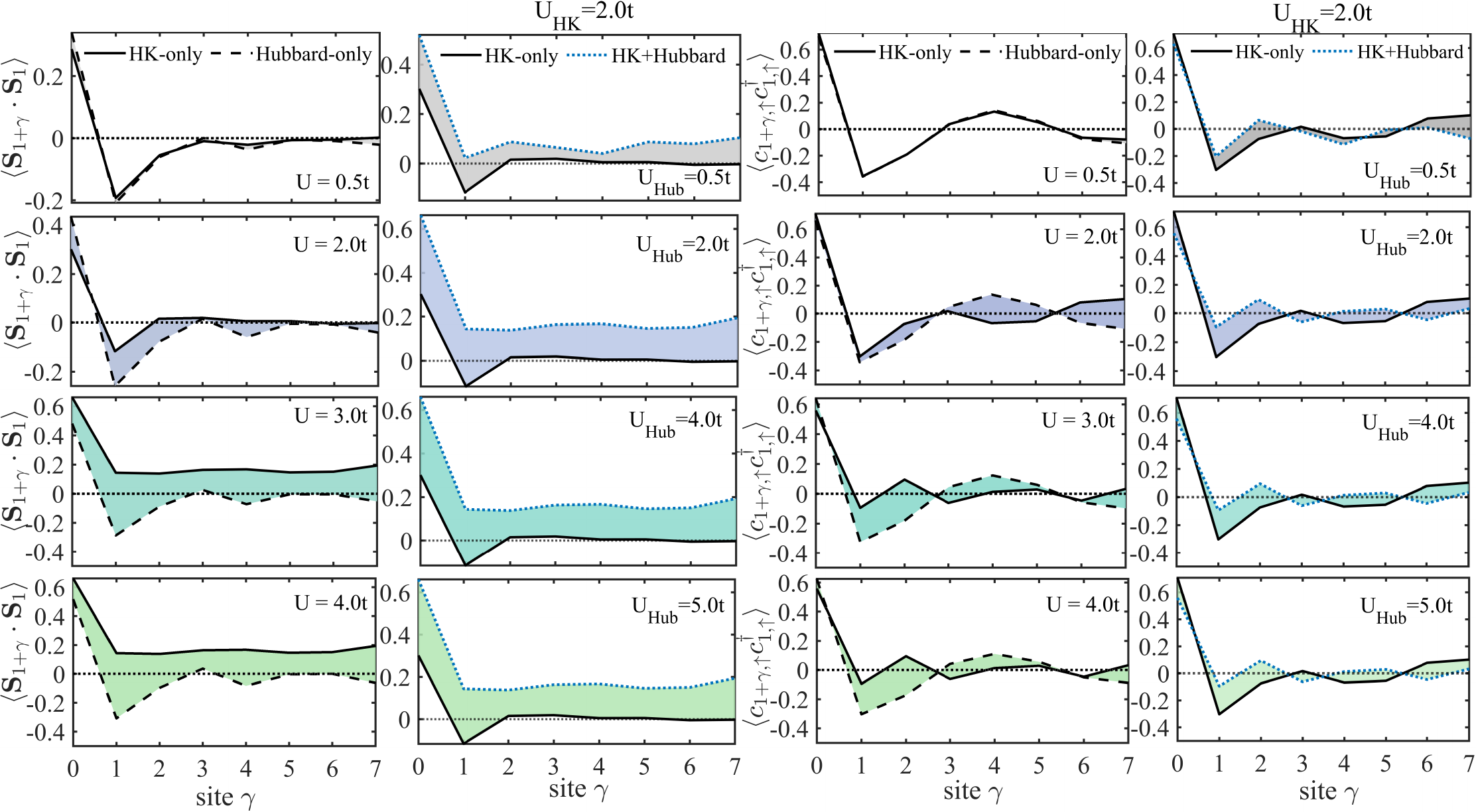}
\caption{Spin and single-particle correlations away from half filling for an $L_x=8$ open chain with $N_\uparrow=N_\downarrow=3$. The left two columns show the spin correlation $C_S(\gamma)$, while the right two columns show the single-particle correlation $C_c(\gamma)$. In the first and third columns, the HK-only and Hubbard-only models are compared at the same interaction strength $U$, with $U_{\rm HK}=U$ and $U_{\rm Hub}=U$, respectively. In the second and fourth columns, $U_{\rm HK}=2t$ is fixed and $U_{\rm Hub}$ is varied in the HK+Hubbard model.}
\label{fig20}
\end{figure*}

\begin{figure*}[t]
\vspace{1.5em}
\centering
\includegraphics[width=0.8\textwidth]{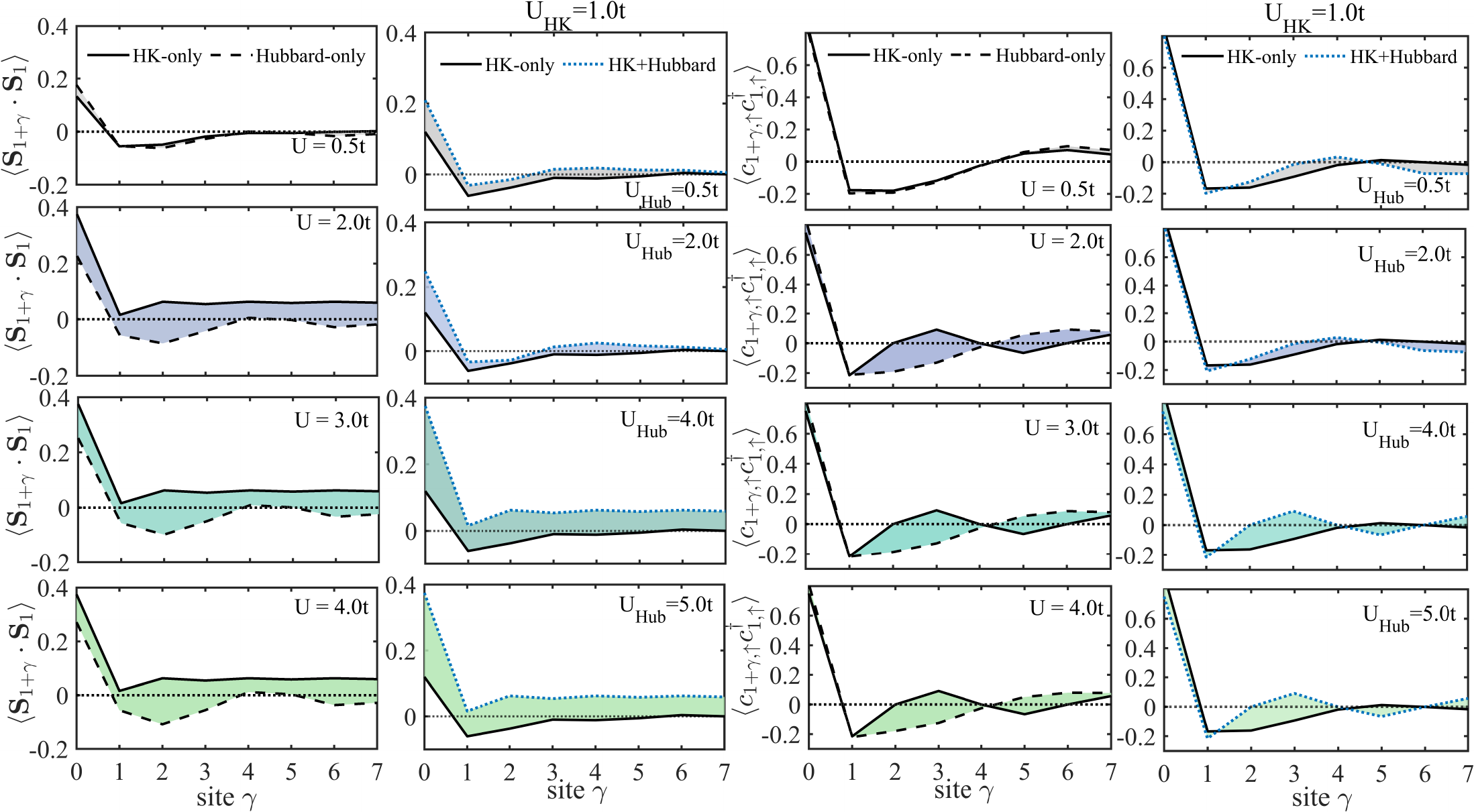}
\caption{Same as Fig.~\ref{fig20}, but for $N_\uparrow=N_\downarrow=2$. Here the second and fourth columns are obtained with fixed $U_{\rm HK}=1.0t$.}
\label{fig21}
\end{figure*}

To examine whether the Hubbard-assisted HK tendency survives away from half filling, we calculate the spin and single-particle correlations for two doped fillings. We consider the same $L_x=8$ open chain in the $S^z=0$ sector with $N_\uparrow=N_\downarrow=3$ and $N_\uparrow=N_\downarrow=2$, corresponding to total particle numbers $N=6$ and $N=4$, respectively.

Figures~\ref{fig20} and \ref{fig21} show that doping modifies the detailed correlation profiles compared with the half-filled case. In particular, the spin correlations no longer develop the nearly distance-independent positive background observed in the strong-HK regime at half filling, and the single-particle correlations retain finite long-distance components. Nevertheless, the characteristic Hubbard-assisted HK tendency remains visible. For fixed $U_{\rm HK}=2t$ and $1t$, increasing $U_{\rm Hub}$ continuously shifts both the spin and single-particle correlations toward the corresponding HK-only results with stronger $U_{\rm HK}$. The spin correlations evolve toward the more positive HK-like pattern, while the single-particle correlations become increasingly suppressed relative to the weak-HK limit.

These results show that although doping changes the detailed form of the correlation functions, the onsite Hubbard interaction still enhances the HK-type correlation tendency over a finite doping range.

\FloatBarrier
\bibliographystyle{apsrev4-2}
\bibliography{Refs}

\end{document}